\documentclass[sigconf, nonacm]{acmart}

\newcommand\vldbpagestyle{plain}

\newcommand{\paperTitleFull}{Computing in the Era of Large Generative Models: \\From Cloud-Native to AI-Native}

\usepackage{anyfontsize}
\usepackage[hyphens]{url}

\definecolor{linkcolor}{HTML}{647382}
\definecolor{citecolor}{HTML}{647382} %
\definecolor{urlcolor}{rgb}{0.4,0.2,0.2}
\definecolor{sqlcolor}{HTML}{965d67}
\definecolor{smtcolor}{HTML}{5d968c}
\definecolor{webblue}{rgb}{0,0,.7}
\definecolor{webgreen}{rgb}{0,.5,0}
\definecolor{webbrown}{rgb}{.6,0,0}
\usepackage{array}
\usepackage{tikz}
\usepackage{frame}
\usepackage{mathtools}
\usepackage{mdwlist}
\usepackage{listings}
\usepackage{amsmath}
\usepackage{float}
\usepackage[short]{optidef}
\usepackage{algpseudocode}
\usepackage{color}
\usepackage{hyperref}
\usepackage{multirow,mathrsfs}

\usepackage{amsmath}
\usepackage{subcaption}

\usepackage{endnotes,microtype,xspace,graphicx,fancyvrb,multirow}
\usepackage{supertabular,booktabs}
\usepackage{array,relsize}
\usepackage{fancyhdr}
\usepackage{enumitem}
\usepackage{balance}
\usepackage{booktabs}
\usepackage{pifont}
\usepackage{listings}
\usepackage{multirow}
\usepackage[scaled]{beramono}
\usepackage{tabularx}
\makeatletter
\newcommand\BeraMonottfamily{%
  \def\fvm@Scale{0.85}
  \fontfamily{fvm}\selectfont
}
\makeatother

\usepackage{aliascnt}

\usepackage[ruled,linesnumbered]{algorithm2e}
\usepackage{setspace}
\usepackage{enumitem}
\usepackage{ hyperref }
\usepackage{dsfont}
\usepackage{svg}

\SetAlFnt{\small}
\SetAlCapFnt{\small}
\SetAlCapNameFnt{\small}

\usepackage{semantic}
 
\usepackage{stmaryrd}
\usepackage{ltablex}
\usepackage{mathtools}
\usepackage{adjustbox}
\usepackage[capitalize,noabbrev,nameinlink]{cleveref}

\crefname{lstlisting}{listing}{listings}
\Crefname{lstlisting}{Listing}{Listings}

\newcommand{\cut}[1]{}

\title{\paperTitleFull} 
\author{Yao Lu$^{1,2}$, 
Song Bian$^3$,
Lequn Chen$^4$,
Yongjun He$^5$,
Yulong Hui$^{6}$,
Matthew Lentz$^{13}$,\\
Beibin Li$^{19}$,
Fei Liu$^{18}$,
Jialin Li$^2$,
Qi Liu$^{7}$,
Rui Liu$^8$,
Xiaoxuan Liu$^9$,
Lin Ma$^{10}$,\\
Kexin Rong$^{11}$,
Jianguo Wang$^{12}$,
Yingjun Wu$^{17}$,
Yongji Wu$^{13}$,
Huanchen Zhang$^{6}$,\\
Minjia Zhang$^{14}$,
Qizhen Zhang$^{15}$,
Tianyi Zhou$^{16}$,
Danyang Zhuo$^{13}$}

\affiliation{%
  \institution{\vspace{0.05in}Nativ$^1$, National University of Singapore$^2$, 
  University of Wisconsin, Madison$^3$, 
  University of Washington$^4$, 
  ETH Z\"urich$^5$, 
  Tsinghua University$^{6}$, 
  University of Hong Kong$^{7}$, 
  University of Chicago$^8$, 
  University of California, Berkeley$^9$, 
  University of Michigan$^{10}$, 
  Georgia Institute of Technology$^{11}$,  
  Purdue University$^{12}$, 
  Duke University$^{13}$, 
  University of Illinois, Urbana-Champaign$^{14}$, 
  University of Toronto$^{15}$, 
  University of Maryland$^{16}$, 
  RisingWave$^{17}$,
  ByteDance$^{18}$,
  Microsoft$^{19}$\vspace{0.05in}}
}

\begin{document}

\newcommand{\mail}[1]{\href{mailto:#1}{#1}}

\begin{abstract}
In this paper, we investigate the intersection of large generative AI models and cloud-native computing architectures. Recent large models such as ChatGPT, while revolutionary in their capabilities, face challenges like escalating costs and demand for high-end GPUs.  Drawing analogies between large-model-as-a-service (LMaaS) and cloud database-as-a-service (DBaaS), we describe an AI-native computing paradigm that harnesses the power of both cloud-native technologies (e.g., multi-tenancy and serverless computing) and advanced machine learning runtime (e.g., batched LoRA inference). These joint efforts aim to optimize costs-of-goods-sold (COGS) and improve resource accessibility. The journey of merging these two domains is just at the beginning and we hope to stimulate future research and development in this area. 
\end{abstract}

\maketitle 

\pagestyle{\vldbpagestyle}
\begingroup
\endgroup


\section{Introduction}
Recent advancements have witnessed the advent of generative AI, notably in the realm of stable diffusion models such as SDXL and large language models such as ChatGPT. These models have shown revolutionary abilities in understanding human language and generating realistic content. As these models evolve and increase in size and complexity, their applications expand across a wide range of domains, from advanced chatbots and virtual assistants to data analysis, forecasting, and even creative endeavors like writing, art, and music composition.

However, the proliferation of these models and their applications has brought about unique challenges to the systems and infrastructures underneath, both vertically in terms of increased individual model and task complexity, and horizontally with the exploding number of users and applications. Among many practical concerns, \emph{cost of goods sold (COGS)} is one of the major barriers to pushing these models further in AI applications. 
Take, for instance, OpenAI's GPT-4. The model inference is priced at \$0.12 per 1K tokens\footnote{With 32K context. Base model inference price in October 2023. }, which is more than the expected revenue per usage in many web-based applications such as search engines \cite{costreport}. The business justification for using these large models comes under scrutiny. To train these models often at the scale of tens to hundreds of billions of parameters, the vast amounts of data required, coupled with the computational power needed, resulted in significant expenses. The iterative nature of model refinement also piles up the costs over time. 
Besides COGS, the \emph{accessability} of GPUs, essential for training and serving large models, has also emerged as a bottleneck. The surging demand for large models has intensified the competition for high-end GPUs; many research institutions and small enterprises grapple with limited access to these specialized hardware resources. 

In comparison, models such as Meta's Llama-2~\cite{touvron2023llama}, offer a notable financial benefit due to their size; being smaller (a few to tens of billions of parameters), these models are inherently cheaper to train, use, and maintain. Instead of aiming for broad capabilities as seen in gigantic generic AI models, these models are often fine-tuned for specific tasks. This specialization ensures optimized performance without the overhead of unnecessary model abilities~\cite{hestness2017deep,wei2022emergent,alabdulmohsin2022revisiting}. The large number of specialized models adds to the complexity of the systems underneath.  

The current AI software stack features a modular architecture. Model runtime frameworks such as TensorFlow~\cite{abadi2016tensorflow} and PyTorch~\cite{paszke2019pytorch} are the key enablers for upper-layer systems such as Huggingface Transformers~\cite{wolf2019huggingface}, Megatron-LM~\cite{shoeybi2019megatron, narayanan2021efficient} and DeepSpeed~\cite{rasley2020deepspeed, smith2022using, rajbhandari2022deepspeed}, amplifying the model training and deployment efficacy in distributed settings. However, current efforts predominantly concentrate on vertical scalability and efficiency in single-model systems. The challenges associated with resource accessibility and optimizing COGS in extensive \emph{scale-out configurations} have yet to become research and development focuses.

\vspace{0.05in}
\noindent\textbf{Looking back: cloud-native technologies.} 
Arguably, large generative models, at a high level, function analogously to databases. They capture data knowledge and, upon receiving a query or prompt, assemble and provide a relevant response. 
To enhance the COGS and resource accessability for large generative models, it is instructive to revisit the evolution of the cloud as well as database-as-a-service (DBaaS) over recent years. Importantly, the challenges faced are not unprecedented; many fall under the purview of cloud-native computing, a paradigm shift that has redefined our understanding of the cloud ecosystem.  

Specifically, by utilizing containers and orchestrators like Kubernetes~\cite{kubernetes}, cloud-native computing ensures scalability, resilience, and modularity. The microservices structure decomposes systems into distinct modules, enhancing agility and simplifying maintenance. Multi-tenancy permits multiple systems and tasks to share common infrastructures, thus optimizing resources and cost while ensuring isolation. With auto-scaling and serverless computing, the cloud-native architecture significantly improves COGS when usage patterns change~\cite{jonas2017occupy, jonas2019cloud}.
In this vision paper, we unravel the nuances of how systems for large generative models can be seamlessly woven into the cloud-native computing architecture. 

\vspace{0.05in}
\noindent\textbf{AI-native: new challenges and opportunities.} Inherit from the legacy cloud-native systems, several of the aforementioned technologies can be effectively applied, including the containerization of machine learning runtime and the dynamic scaling of model inference tasks. It is worth noting that, in these scenarios, machine learning operations are often treated as black boxes. By fostering a deeper integration and potentially co-designing machine learning runtime with cloud-native systems, we pave the way for the emergence of a novel \emph{AI-native} computing paradigm.

At the core of the AI-native paradigm lies the process of training, fine-tuning, and deploying large models, with goals remaining unchanged for improved COGS and resource accessability. 
Take the following use case for instance of AI-native computing: as a majority of end users cannot afford to train foundational models, they fine-tune open-source models with their specific data and applications. Currently serving hundreds of models at the same time can be of low efficiency, even if these models are fine-tuned from the same foundational model. In fact, this is a typical multi-tenancy scenario. We note a prior work, Punica~\cite{punica}, that developed an ML runtime with a batched LoRA inference mechanism which improved the output throughput by up to 14x. This is an exemplar of the AI-native computing paradigm.

To alleviate the resource accessibility issue, we may turn to serverless computing and more versatile cloud infrastructures from emerging decentralized GPU providers such as Vast.ai and Akash Networks who offer cost-effective GPU containers. However, hosting diverse large-model jobs and building, operating, and optimizing systems upon such heterogeneous and sometimes geo-distributed infrastructures create unforeseen challenges.

Nevertheless, COGS and resource accessability are merely the tip of the iceberg for this AI-native paradigm. Equally vital topics include serving broader AI applications and so on; we defer more discussions to Section \ref{sec4}. These challenges highlight the complexities involved in the deep integration of cloud-native methodologies with large generative models and suggest numerous exciting opportunities for future research and development.

\vspace{0.05in}
\noindent\textbf{Key takeaways} of this paper can be summarized as follows:  
\begin{itemize}[leftmargin=0.2in]
\item The current AI software stack focuses on vertical scalability and efficiency in single-model systems; horizontal scaling-out scenarios have yet to become research and production focuses. 
\item Large-model-as-a-service (LMaaS) and database-as-a-service (DBaaS) have commonalities. Many cloud-native designs are readily applicable to improve system efficiency and other aspects. Meanwhile, many other techniques require further adaptations and co-design between ML and cloud systems. This formulates a novel AI-native computing paradigm, and we provide a tentative list of interesting topics
\item Many open questions remain. We provide an outlook on the research and production challenges and opportunities. 
\end{itemize}


\section{Background} \label{sec2}
\noindent\textbf{Cloud-native computing} has redefined software development, leveraging concepts such as containerization and microservices for scalability and adaptability. Classic software like legacy customer relationship management (CRM) platforms and database systems have been limited by the constraints of proprietary infrastructures. One cornerstone of the transformation towards cloud-native computing is containerization which allows developers to wrap applications in containers, ensuring uniform behavior in heterogeneous environments. Kubernetes~\cite{kubernetes} and other container orchestration tools take this a step further. Say an e-commerce platform witnesses a sudden surge in traffic during a sale, an orchestrator can dynamically scale resources, ensuring that the website remains responsive.

The microservice architecture decomposes applications into discrete, function-oriented services. Cloud-native platforms like YouTube exemplify this strategy. Instead of a monolithic structure, they operate via microservices, each dedicated to specific tasks, ranging from user authentication and data storage to video encoding and streaming~\cite{gan2019open}. This approach offers several advantages. First, it enhances robustness; isolated failures in one service will not disrupt others. Second, maintenance becomes more manageable; individual services can be updated or debugged without affecting the entire ecosystem. Last, this architecture improves agility. Developers can concurrently innovate and roll out new features.

It is worth noting that multi-tenancy plays an important role in cloud-native computing. Rooted from the shared-everything architecture in the 80s, software platforms started to harness multi-tenancy to serve multiple users with shared resources on the set of infrastructures, instead of spinning up individual instances for each user~\cite{hindman2011mesos, ghodsi2011dominant, mahajan2020themis, narayanan2020heterogeneity}. This greatly optimizes resource usage and reduces user costs. Amongst many successful cloud-native systems, database-as-a-service (DBaaS)  exploits multi-tenancy from various aspects including infrastructures,  datasets, queries, and relational operators. Prominent industry players like Snowflake~\cite{dageville2016snowflake} and Amazon Redshift~\cite{gupta2015amazon} have emerged, driven by the imperative to provide enhanced services at competitive price points.

\vspace{0.05in}
\noindent\textbf{Large-model-as-a-service} (LMaaS),  fundamentally a cloud-based approach, bridges the gap between advanced AI and real-world applications by offering efficient, scalable, and accessible deployment mechanisms. Organizations, especially those without vast computational resources, can access state-of-the-art models without the overhead of training, maintaining, and deploying the models. 
By abstracting the complexities from the systems and infrastructures aspects, LMaaS offers simple interfaces to end users: 

\vspace{0.05in}
\emph{Model training.} $\mathcal{W}_m\leftarrow\texttt{train}(m,\mathcal{D},\mathcal{C})$ builds the model $m$ on dataset $\mathcal{D}$ with hyper-parameters $\mathcal{C}$; the output is the weight (i.e., model parameters) of the model $\mathcal{W}_{m}$.     
    
\vspace{0.03in}
\emph{Model fine-tuning}. $\mathcal{W}_m'\leftarrow\texttt{finetune}(m, \mathcal{W}_m, \mathcal{D}', \mathcal{C}')$  is a special case of model training to gain domain-specific knowledge at a much smaller cost. The input includes the weights $\mathcal{W}_m$ of the previously-trained model. Different training hyper-parameters $\mathcal{C}'$ are used with a dataset $\mathcal{D}'$ that is also often smaller. In a common case, the disparity in $\mathcal{C}'$ includes a much smaller number of epochs while most other knobs remain unchanged.
    
\vspace{0.03in}
\emph{Model inference}. $r\leftarrow\texttt{inference}(m,\mathcal{W}_m, q)$ generates content {$r$} with a trained model and an input query $q$. 

\vspace{0.05in}
Recently, Low Rank Adaptation (LoRA)~\cite{hu2021lora},  a fine-tuning strategy to further reduce computational cost, has attracted both research and production attention. Such a strategy only slightly changes the user contract:
for model fine-tuning, the contract changes to $\Delta m,\mathcal{W}_{\Delta m}\leftarrow\texttt{finetune}(m, \mathcal{W}_m, \mathcal{D}', \mathcal{C}'')$. The key idea is to keep the weights of the base model $\mathcal{W}_m$ unchanged while training a small \emph{LoRA adaptor} $\mathcal{W}_{\Delta m}$ that has only small fractional trainable parameters. As a result, the fine-tuning speed is faster by orders of magnitude. 
Therefore, the new model inference contract is $r\leftarrow\texttt{inference}([m,\Delta m],[\mathcal{W}_m,\mathcal{W}_{\Delta m}], q)$  where the input query $q$ goes through both the base and the adaptor models; doing so 
incurs little impact on the inference latency since the adaptor model is small.

LoRA provides an agile and low-cost contract in addition to the simple LMaaS interfaces. The base model can be viewed as a public asset for many different users and application scenarios, while the adaptors serve as specifications. We will show how this new contract plays an important role in cloud-native, AI-native computing for the remainder of this paper. 

HuggingFace and similar marketspaces are yet another driving horse of LMaaS that provides the community with state-of-the-art pre-trained models, finetuned adaptors, datasets, and a platform for training and deploying custom models. Thousands of established organizations publish models and other resources on their platform which no doubt pushes the entire ecosystem of LMaaS further.

\section{Cloud-native for large models}\label{sec3} 
The intersection of LMaaS and cloud-native computing is not merely a convergence of two technological domains; it is a fusion that has the potential to reshape the very fabric of modern cloud and AI. The synergies can be described below.

\vspace{0.05in}
\emph{Containerization}. It is worth noting that recent advancements in virtualization technologies enable a transparent pass-through to low-level operating systems. On the AI and deep learning framework level, popular offerings such as PyTorch and TensorFlow have integrated the runtime for heterogeneous compute architectures such as CPU, GPU and FPGA. These developments, as well as the protocol design for collaborative work in-between various container instances, facilitate effortless delivery of the software stack for large models in dynamic deployment scenarios. 

\vspace{0.03in}
\emph{Orchestration} plays a critical role in ensuring elasticity, scalability, and resilience, which are fundamental for LMaaS to deliver robust and economically viable pay-as-use services. Given the fluctuating nature of user workloads and the underlying infrastructure dynamics, it is imperative to draw from, and further refine, the best practices inherent in current cloud-native solutions. 
Take, for instance, the changing query patterns: as they shift from predictable or sporadic requests to intense traffic surges, serverless technologies are equipped to dynamically and quickly scale resources upwards and downwards. They can also facilitate the migration of containers tasked with model-serving functionalities.
Such adaptability 
minimizes resource redundancy and thus improves the COGS.
Yet another dimension to consider is the disparate resource availability and heterogeneity across various segments of the cloud. Ensuring efficient orchestration in such a diverse environment demands meticulous resource management strategies involving a mixture of resource allocation, queuing, and adjustments on-the-fly, especially when catering to large-scale, distributed ML workloads.

\vspace{0.03in}
\emph{Microservices}.  Beyond the fundamental services previously mentioned, there is a clear shift towards 
integrating large models with enhanced functionalities and additional data.
As the machine learning lifecycle matures, it increasingly intersects with sophisticated algorithms and methodologies. A case in point is the recent adoption of multi-stage training strategies in large language models, particularly the use of Reinforcement Learning with Human Feedback (RLHF)~\cite{ouyang2022training}. Concurrently, the advent of diverse AI applications, including search engines, speech synthesis, business data analytics, and 
image generation, demands processing capabilities for multi-modal data. Given these complexities, cloud providers are compelled to 
architect and streamline microservices and augment the models backed by disaggregated, large-scale data stores.
This is essential to 
accommodate the rapid-growing diversity and volume of data and the evolving landscape of machine learning algorithms.

\vspace{0.03in}
\emph{Multi-tenancy}, a design principle that affects each of the technologies above, can be viewed from different aspects. From the model perspective, there can be multiple base models and each company with multiple LoRA adaptor models; there can also be concurrent input queries from different users. From the task perspective, there can simultaneously be training, fine-tuning, serving, data preparation, management, and operational workloads. In these scenarios, it is essential to consolidate the computation and save GPU memory. These LM use cases also open up opportunities for multi-tenant systems and infrastructures underneath to further save costs and improve efficiency, for example, aggregated network communication and shared data infrastructures in distributed training and dataset preparation tasks.  

\vspace{0.05in}
Armed with these capabilities, we are poised to establish cloud-native infrastructures tailored for large generative models. In the subsequent sections, we will describe our initial endeavors within select areas of the aforementioned research spectrum.

\vspace{0.05in}
\noindent\textbf{Example: RAG-as-a-Service vs BI-as-a-Service}. Retrieval Augmented Generation (RAG)~\cite{lewis2020retrieval} has become an important topic in natural language processing. It extends the capabilities of large language models by incorporating an information retrieval system for reference data. RAG is versatile in diverse applications such as document analysis, summarization, personal chatbot interactions, and code comprehension.

\begin{figure}[t]
\centering
\includegraphics[width=1\linewidth]{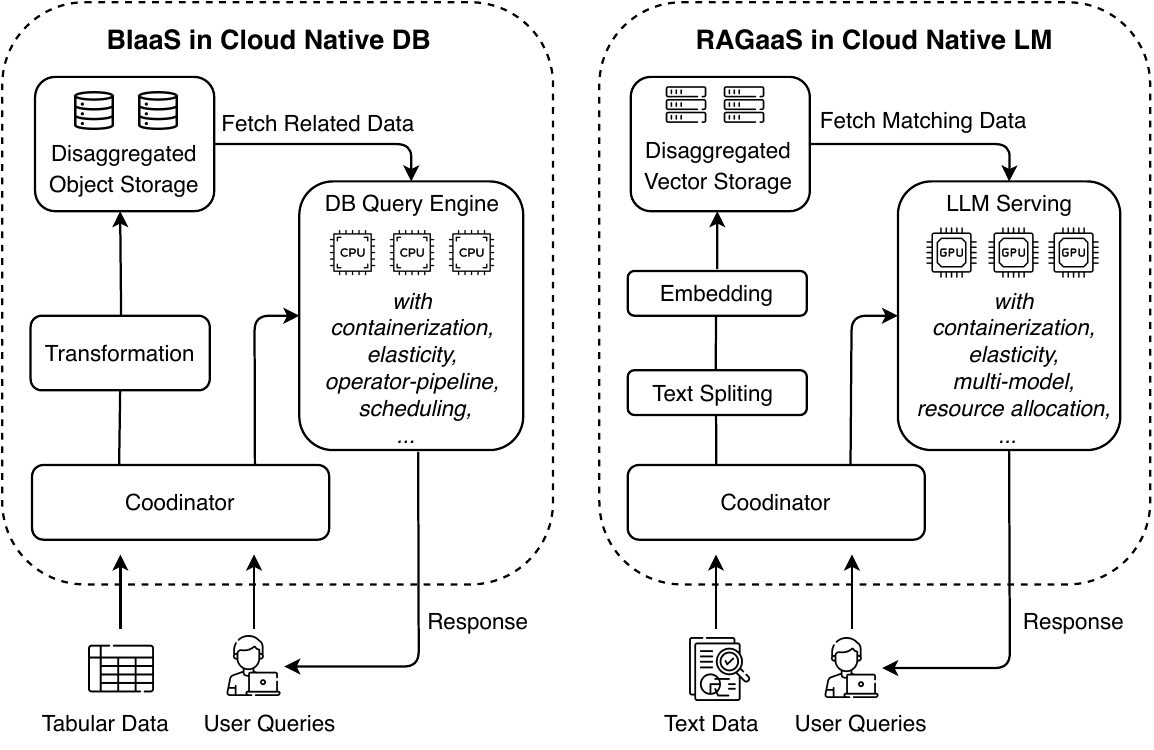}
\caption{BIaaS and RAGaaS share architectural commonalities. }
\label{fig:q1}
\end{figure}

Figure 1 (Right) illustrates the workflow and architecture of RAG-as-a-service (RAGaaS), compared with a typical cloud-native Buiness-Intelligence-as-a-Service (BIaaS)~\cite{chaudhuri2011overview} architecture in Figure~\ref{fig:q1} (left). In the right figure, the process begins with the extraction of text documents or code from clients, followed by segmenting inputs into smaller units. These segments are then transformed into vectors and stored in a dedicated vector storage system. When a user submits a query, the system retrieves the most relevant data chunks from the vector storage. The user's query and the retrieved data are then combined and passed into the LM inference model to generate an answer.
In comparison, BIaaS in contemporary cloud-native database systems exhibits a similar workflow. It also involves data processing stages for extracting, splitting, embedding, and loading tabular input data. SQL queries initiate data retrieval and execution of corresponding database operations, mirroring the data retrieval and LM inference processes in RAG.

\vspace{0.05in}
\noindent\textbf{Adopting cloud-native techniques}. When bridging the concepts of database and large model applications, some cloud-native technologies can be readily applied in the context of RAGaaS. For instance, large model inference endpoints can be containerized for streamlined deployment; functionalities like vector database search can be offered as microservices. A cloud orchestrator improves system efficiency by leveraging disaggregated storage and dynamically scaling containers and GPUs resources. To enhance multi-tenant RAG workload processing, implementing query optimization and batching techniques have proven to be effective, enabling batched concurrent queries and routing different queries to various large models in accordance with user service level agreements.

While cloud-native technologies offer immediate benefits, treating large model tasks and system components as black boxes, there remains a wealth of untapped potential for specialized large-model optimizations. 

\section{Preliminary AI-native computing} \label{sec4}
In this section, we explore the prospects arising from the deep integration of large models with cloud-native technologies, paving the way for a novel AI-native computing paradigm. To achieve so, we mention some prior work as well as some of our preliminary experiments. Our purpose here is not for completeness but to point out the potential values for future research in this area. 

\begin{figure}[t]
\begin{subfigure}[b]{.495\linewidth}
\centering
\includegraphics[width=\linewidth]{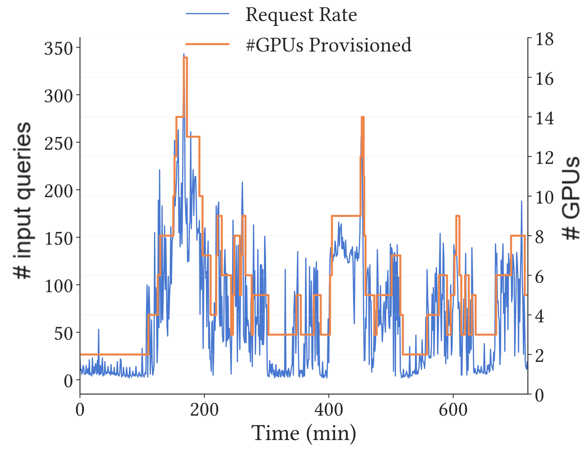}
\caption{Elastic resource scheduling.}
\label{fig:serverless}
\end{subfigure}
\hfill
\begin{subfigure}[b]{.495\linewidth}
\centering
\includegraphics[width=\linewidth]{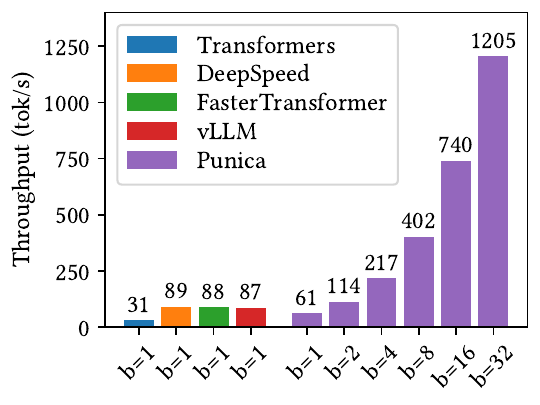}
\caption{Batched LoRA inference. }
\label{fig:multi-lora}
\end{subfigure}
\caption{Preliminary results to verify AI-native computing.}
\end{figure}

\vspace{0.05in}
\noindent\textbf{Case Study 1: Elasticity}. The first idea from legacy cloud-native computing that we try out is the elastic resource scheduling for serving 7B llama-2 models on a cluster of NVidia 4090 GPUs. This is of practical importance to improve resource utilization when incoming query patterns vary. Our cloud orchestrator collects runtime telemetries and applies a heuristics-based feedback mechanism to scale up/down the number of inference instances. In our preliminary experiments, we synthesize incoming queries with a realistic whole-day pattern and demonstrate our findings in \autoref{fig:serverless}. Surprisingly, with simple scaling heuristics, our system already provides prompt scaling; models can be loaded quickly in seconds, with the aid of high-speed networking and PCI-E 5.0 bus. This indicates that elasticity is possible and beneficial in the new AI-native computing paradigm to improve resource usage efficiency. 

\vspace{0.05in}
\noindent\textbf{Case Study 2: Multi-tenancy}.  Here we elaborate more on the example shown in the earlier sections. When many LoRA fine-tuned models~\cite{hu2021lora} share the same foundational model, a novel machine learning runtime, Punica~\cite{punica}, allows batching inference jobs from multiple user queries. This is a typical multi-tenant scenario in classic cloud systems where users share the same software infrastructures. The delta here is the model inference procedure which involves one single copy of the base model and multiple LoRA adaptors. Punica implement a LoRA batching mechanism with a custom CUDA kernel which enables fast communication between the base model and adaptors. 
\autoref{fig:multi-lora} compares Punica with state-of-the-art large model serving systems, with b = batch size on the x-axis, and baseline systems such as Huggingface Transformers~\cite{wolf2019huggingface}, DeepSpeed~\cite{rasley2020deepspeed}, and vLLM~\cite{kwon2023efficient} running at batch size 1, since they do not support batched LoRA inference. Punica delivers 14x throughput compared to vLLM at a batch size of up to 32 with 7B llama-2 base model on a single A100 GPU with 80G memory.

\vspace{0.05in}
\noindent\textbf{Case Study 3: Hybrid cloud deployment and optimization}. As the cloud paradigm expands and becomes increasingly diverse, notably by the ubiquitous smart devices and the Internet of Things. A prior work, JellyBean~\cite{wu2022serving}, introduced a system that deploys and optimizes machine learning inference workloads across such heterogeneous infrastructures. 

JellyBean has the following cloud- and AI-native designs: (1) containerized runtime for deploying ML inference jobs on heterogeneous infrastructures such as IoT devices and cloud data centers, and (2) a hybrid cloud orchestrator that imploys query optimization techniques to choose the most cost-effective models aligned with specified service-level objectives (like throughput or accuracy) and assigns workers across different segments of the cloud.
With these designs, JellyBean empirically reduces the total serving cost by up to 58\% for tasks like visual question answering and by up to 36\% for vehicle tracking from the NVIDIA AI City Challenge, compared to prior solutions for hybrid-cloud ML deployment. 

\vspace{0.05in}
\noindent\textbf{Remark}.
Through three case studies, the advantages of a cloud- and AI-native design in enhancing the efficiency and mitigating the costs of large models become evident. While our investigations remain preliminary and address only specific aspects, we anticipate comprehensive and in-depth studies to follow, given the rising significance of large generative models in research and production.


\section{Outlook to the future} \label{sec5}
While the previous section exemplifies a few valuable directions in the AI-native computing paradigm, there are more open challenges and opportunities that are equally important. We elaborate on some of them in this section.  

\vspace{0.05in}
\noindent\textbf{Runtime for large generative models.} 
While this vision paper mostly discusses scaling large generative models horizontally, vertical scalability and efficiency of the machine learning runtime are still of critical importance. Large models require substantial memory that often exceeds the GPU capacity; key-value caches used in the transformers push this further. Transferring data between High-Bandwidth Memory (HBM) and Static Random-Access Memory (SRAM) consumes memory bandwidth, hence sparse computations make inference bandwidth-bound and result in underutilized GPU compute capacities~\cite{sheng2023flexgen}. Building an efficient inference engine that fully leverages the available memory and computation remains a prevailing challenge. 

Recent advancements in machine learning have introduced the concept of Mixture of Experts (MoE) models~\cite{shazeer2017outrageously, fedus2022switch} and speculative decoding~\cite{chen2023accelerating, leviathan2023fast} to enhance the model's efficiency and quality. The primary idea behind these innovations is to curate a collection of diverse, smaller models, each tailored for specific tasks, thus avoiding the exhaustive use of complete models and conserving computational resources.  For example, models like Llama-2 already exhibit varying sizes, making them more cost-effective when used strategically. 
However, deciding if the potential savings and increased efficiency outweigh the initial overhead of creating and maintaining multiple models is non-trivial. Insights from database management systems, such as materialized views~\cite{goldstein2001optimizing}, can be applied to efficiently handle recurring queries without repetitively calling large models.

As spot innovations continue to advance rapidly, we anticipate that significant breakthroughs will emerge from both the systems and machine learning domains. Examples include FlashAttention~\cite{dao2022flashattention, dao2023flashattention} and PagedAttention~\cite{kwon2023efficient} mechanism which incorporates ideas from both domains; these innovations will push current solutions to the next level.

\vspace{0.05in}
\noindent\textbf{Continuous learning and serving systems.} In many applications, data comes in real-time. To provide users with useful and up-to-date insights, serving systems may be required to keep learning from data streams. Traditional approaches of periodically re-training or updating large models prove to be computationally expensive and time-prohibitive~\cite{li2022warper}. 

Systems that learn continuously are the solution to keep the models afresh. Classic stream processing systems like Naiad~\cite{murray2013naiad}, Spark Streaming~\cite{zaharia2013discretized, venkataraman2017drizzle}, Apache Flink~\cite{carbone2015apache}, and RisingWave can be adapted to process continuous data streams in (near) real-time. Machine learning and data management technologies such as LoRA fine-tuning~\cite{hu2021lora} and indexing can be used to process new data and update existing models efficiently. Machine unlearning~\cite{bourtoule2021machine}, i.e., forgetting specific knowledge, is notable in future research and production challenges to remove data that is not needed. 

Real-time serving of large models presents challenges, including scenarios where queries arrive in parallel but are not synchronized~\cite{gujarati2020serving, shen2019nexus, pope2023efficiently}. Prior research such as Orca~\cite{yu2022orca} has provided tentative solutions. The complexity increases in multi-tenant systems where many base models, users, applications, and tasks compete for the same resources. We anticipate future research will systematically advance solutions in this domain.

\vspace{0.05in}
\noindent\textbf{Service availability.}  
As large generative models become essential infrastructure services, ensuring their resilience to failures is crucial. Unlike database services that incorporate logging, checkpointing, and replication in cloud-native architectures for both recoverability and availability, large model training on distributed GPU servers is resource-intensive and time-consuming~\cite{zhang2022opt}. Handling partial resource failures is vital to maintaining model performance~\cite{wang2023gemini}. Straw-man solutions like periodic checkpointing are costly and lead to GPU underutilization. Asynchronous approaches with fine-grained logging offer a more promising direction for fault-tolerant training, allowing for progress persistence without halting GPU execution.
In contrast, online inference prioritizes availability over recoverability, as tasks are short and can be re-executed. Replicating inference deployments across multiple fault domains eliminates downtime but increases costs. Cloud-native proposals suggest cost optimizations through multi-tenancy, enabling replication for higher throughput and graceful failure recovery.

High availability is a challenge in LMaaS due to heterogeneous hardware, fragile machine learning software, and evolving user and application requirements. Future research opportunities include designing availability-guaranteed replication and recovery algorithms, hardware-software co-design for faster crash recovery, leveraging cloud storage architectures with specialized caching, and so on. Addressing these challenges is crucial as large language models become integral to various applications.

\vspace{0.05in}
\noindent\textbf{Resource accessability and ephemerality.}  Training and deploying large generative models often requires a multitude of GPUs, especially during foundational model pre-training. High-bandwidth networking is crucial for inter-node communication and gradient synchronization. However, this gives rise to resource accessibility issues, making it challenging to access such infrastructures. To address this, previous research has focused on training large models using geo-distributed infrastructures~\cite{yuan2022decentralized}, and there are high expectations for further developments in this field. Simultaneously, cloud providers are also moving towards disaggregation. Global GPU market spaces like Vasi.ai and Akash Networks offer cost-effective rentals with highly heterogeneous compute nodes distributed worldwide. However, constructing cloud services on such hyper-disaggregated infrastructures poses significant challenges in communication, fault tolerance, and resource management. Nevertheless, the potential benefits are substantial.

Ephemeral cloud resources are increasingly prevalent, such as spot instances \cite{amazon-spot2023} where short-lived computational resources are provided and the rising zero-carbon clouds~\cite{DBLP:journals/cacm/Chien21a, patterson2021carbon} where the infrastructures rely heavily on sustainable energy sources (e.g., wind power). These resources often have factional prices compared to the rates for regular reserved or on-demand resources \cite{amazon-spot-pricing2023,DBLP:journals/csur/ShafieiKM22} and can exhibit significant fluctuations over time. It is challenging to deploy LM tasks on ephemeral resources. Given that LM training is typically long-running, it is essential to establish an efficient mechanism for model checkpointing and reloading. This mechanism allows training processes to be paused when resources become temporarily inaccessible or when prices become prohibitively high, and then resume at a more opportune time without bringing substantial overheads. LM inferences are often characterized by their sensitivity to latency; timely and reasonable decision-making becomes crucial for determining which inference jobs should be prioritized and given access to available resources.

\vspace{0.05in}
\noindent\textbf{Diverse microservices.} In addition to the basic large model training, fine-tuning and inference services, recent production offerings start to explore broader microservices. For example, the lifecycle of machine learning contributes multiple useful microservices such as data curation, cleaning, transformation and labeling. Major cloud vendors and DBaaS such as Databricks~\cite{Databricks} and Snowflake~\cite{dageville2016snowflake} already have relevant offerings and integration.  A notable microservice for LMaaS that can be distinct to the aforementioned cases is the usage of vector databases~\cite{VecDBSurvey23} like Milvus~\cite{Milvus} and Pinecone~\cite{Pinecone} to cache intermediate results, frequent queries, or embeddings~\cite{ChatGPTLimitations23, mohoney2023high}. RAG-based applications, discussed earlier in this paper, is another important use case. Vector databases must choose the vector indexes (e.g., a graph-based index HNSW~\cite{malkov2018efficient}) to tradeoff between memory consumption, query performance, and accuracy. Building robust, scalable and efficient vector database as microservices to the cloud system is of both research and production value.


\vspace{0.05in}
\noindent\textbf{AI operations (AIops)} is now vital for managing complex cloud infrastructures. It automates tasks like monitoring, troubleshooting, performance tuning, and resource optimization. In cloud-based systems for large generative models, AIops will be a pivotal tool to ensure the system reliability and efficiency. Challenges and opportunities are both significant. For instance, predicting workload completion times and incoming workload patterns becomes crucial for enhancing system performance and minimizing service downtime~\cite{venkataraman2016ernest, peng2018optimus}. Optimizing the model training and inference configurations (e.g., batch sizes, optimizer parameters) based on the specific workload also has a significant impact on the system performance and resource consumption.
Insights from database management systems, again, can be applied to intelligently automate the operation of LMaaS. For example, query forecasting methods~\cite{ma2018query} can help detect workload patterns and completion times. Learning-based knob-tuning techniques~\cite{van2017automatic, alipourfard2017cherrypick} can also help optimize training and inference configurations. However, database and LM workloads have disparate properties, which warrants deeper investigation and adaptation of existing techniques.

\vspace{0.05in}
\noindent\textbf{Emerging applications and workloads.}  Apart from the RAG-based applications mentioned earlier in this paper, there have been emerging applications of large generative models; in addition to model inference calls, they often employ complex workflows and sometimes involve interactions or feedbacks from 3rd-party software such as compilers and search engines. One notable type of application is AI agents~\cite{park2023generative,liu2023agentbench} which require multi-turn interactions between the agent and human, or among multiple agents for web-shopping, robotics, gaming, and other use cases.  Application builders create AI agents from large model inference endpoints~\cite{wu2023autogen, talebirad2023multiagent, Hong2023MetaGPTMP} using different role-defining prompts.  While many cloud-native techniques still apply in these scenarios, there are challenges and opportunities to optimize these novel workflows. For example, multiple AI agents that are fine-tuned with different skill sets from the same foundational model can be seen as another use case of multi-tenancy. Compute graph and query optimization techniques may be applied to improve the execution. We project that application-specific systems and optimizations will surge in the near future.

\section{Related Work}
\label{sec:related}

\vspace{0.05in}
\noindent\textbf{Systems for large models.} PyTorch~\cite{paszke2019pytorch}, TensorFlow~\cite{olston2017tensorflow}, and JAX~\cite{jax2018github} are among the popular deep learning frameworks. PyTorch used a dynamic computation graph, known as the define-by-run paradigm, while TensorFlow was recognized for its static define-and-run approach; they now support both modes. JAX offers composable transformations of Python functions, facilitating advanced optimizations and just-in-time compilation to heterogeneous architectures. 
A layer above, systems that serve complex large models, especially in distributed setups~\cite{huang2019gpipe, narayanan2019pipedream, narayanan2021memory}, come to attention; Megatron-LM~\cite{shoeybi2019megatron, narayanan2021efficient}, DeepSpeed~\cite{rasley2020deepspeed, smith2022using} and Huggingface Transformers~\cite{wolf2019huggingface} are three exemplars in this category that provide easy-to-use, distributed training and deployment support. To continuously serve large models with concurrent users and queries, Orca~\cite{yu2022orca}, TGI~\cite{tgi}, and DeepSpeed-MII~\cite{deepspeedmii} build the model serving layer with gRPC user endpoints. These systems are necessary components of the AI-native computing paradigm.

\vspace{0.05in}
\noindent\textbf{Cloud-native in production.} Major cloud providers offer a diverse range of cloud-native services. These include Database-as-a-Service (DBaaS) solutions like Amazon RDS~\cite{RDS}, Azure SQL Databases~\cite{antonopoulos2019socrates}. Function-as-a-Service (FaaS) platforms like AWS Lambda~\cite{AWSLambda}, Azure Functions~\cite{AzureFunctions}, and Google Cloud Functions~\cite{GoogleCloudFunctions} enable serverless code execution. Managed Kubernetes services such as Amazon EKS~\cite{AmazonEKS}, Google Kubernetes Engine (GKE)~\cite{GKE}, and Azure Kubernetes Service (AKS)~\cite{AKS} enable container orchestration.  AI/ML services, big data and analytics, IoT platforms are among the offerings that cater to various business needs, demonstrating the versatility of cloud-native technologies.

\section{Conclusions}
This paper explores the similarities between Large-Model-as-a-Service (LMaaS) and Database-as-a-Service (DBaaS), uncovering shared characteristics and suggesting that cloud-native technologies can be leveraged to improve generative AI systems. However, creating an AI-native computing paradigm through this fusion demands deep insights and innovations. The paper presents three use cases with initial findings, emphasizing future research challenges and opportunities.

\bibliographystyle{ACM-Reference-Format}
\bibliography{confs_long, reference}


\begin{thebibliography}{88}


\ifx \showCODEN    \undefined \def \showCODEN     #1{\unskip}     \fi
\ifx \showDOI      \undefined \def \showDOI       #1{#1}\fi
\ifx \showISBNx    \undefined \def \showISBNx     #1{\unskip}     \fi
\ifx \showISBNxiii \undefined \def \showISBNxiii  #1{\unskip}     \fi
\ifx \showISSN     \undefined \def \showISSN      #1{\unskip}     \fi
\ifx \showLCCN     \undefined \def \showLCCN      #1{\unskip}     \fi
\ifx \shownote     \undefined \def \shownote      #1{#1}          \fi
\ifx \showarticletitle \undefined \def \showarticletitle #1{#1}   \fi
\ifx \showURL      \undefined \def \showURL       {\relax}        \fi
\providecommand\bibfield[2]{#2}
\providecommand\bibinfo[2]{#2}
\providecommand\natexlab[1]{#1}
\providecommand\showeprint[2][]{arXiv:#2}

\bibitem[\protect\citeauthoryear{??}{Pin}{[n. d.]}]%
        {Pinecone}
 \bibinfo{year}{[n. d.]}\natexlab{}.
\newblock \bibinfo{title}{{Pinecone}}.
\newblock   (\bibinfo{year}{[n. d.]}).
\newblock
\showURL{%
\url{https://www.pinecone.io/}}


\bibitem[\protect\citeauthoryear{??}{cos}{2023}]%
        {costreport}
 \bibinfo{year}{2023}\natexlab{}.
\newblock \bibinfo{title}{How Microsoft is Trying to Lessen Its Addiction to
  OpenAI as AI Costs Soar}.
\newblock
  \bibinfo{howpublished}{\url{https://www.theinformation.com/articles/how-microsoft-is-trying-to-lessen-its-addiction-to-openai-as-ai-costs-soar}}.
    (\bibinfo{year}{2023}).
\newblock


\bibitem[\protect\citeauthoryear{??}{Cha}{2023}]%
        {ChatGPTLimitations23}
 \bibinfo{year}{2023}\natexlab{}.
\newblock \bibinfo{title}{{LLM Limitations
  (\url{https://zilliz.com/use-cases/llm-retrieval-augmented-generation})}}.
\newblock   (\bibinfo{year}{2023}).
\newblock


\bibitem[\protect\citeauthoryear{Abadi, Barham, Chen, Chen, Davis, Dean, Devin,
  Ghemawat, Irving, Isard, et~al\mbox{.}}{Abadi et~al\mbox{.}}{2016}]%
        {abadi2016tensorflow}
\bibfield{author}{\bibinfo{person}{Mart{\'\i}n Abadi}, \bibinfo{person}{Paul
  Barham}, \bibinfo{person}{Jianmin Chen}, \bibinfo{person}{Zhifeng Chen},
  \bibinfo{person}{Andy Davis}, \bibinfo{person}{Jeffrey Dean},
  \bibinfo{person}{Matthieu Devin}, \bibinfo{person}{Sanjay Ghemawat},
  \bibinfo{person}{Geoffrey Irving}, \bibinfo{person}{Michael Isard},
  {et~al\mbox{.}}} \bibinfo{year}{2016}\natexlab{}.
\newblock \showarticletitle{{TensorFlow}: A System for Large-Scale Machine
  Learning}. In \bibinfo{booktitle}{{\em Symposium on Operating Systems Design
  and Implementation (OSDI)}}.
\newblock


\bibitem[\protect\citeauthoryear{Alabdulmohsin, Neyshabur, and
  Zhai}{Alabdulmohsin et~al\mbox{.}}{2022}]%
        {alabdulmohsin2022revisiting}
\bibfield{author}{\bibinfo{person}{Ibrahim~M Alabdulmohsin},
  \bibinfo{person}{Behnam Neyshabur}, {and} \bibinfo{person}{Xiaohua Zhai}.}
  \bibinfo{year}{2022}\natexlab{}.
\newblock \showarticletitle{Revisiting neural scaling laws in language and
  vision}.
\newblock \bibinfo{journal}{{\em Advances in Neural Information Processing
  Systems\/}}  \bibinfo{volume}{35} (\bibinfo{year}{2022}),
  \bibinfo{pages}{22300--22312}.
\newblock


\bibitem[\protect\citeauthoryear{Alipourfard, Liu, Chen, Venkataraman, Yu, and
  Zhang}{Alipourfard et~al\mbox{.}}{2017}]%
        {alipourfard2017cherrypick}
\bibfield{author}{\bibinfo{person}{Omid Alipourfard},
  \bibinfo{person}{Hongqiang~Harry Liu}, \bibinfo{person}{Jianshu Chen},
  \bibinfo{person}{Shivaram Venkataraman}, \bibinfo{person}{Minlan Yu}, {and}
  \bibinfo{person}{Ming Zhang}.} \bibinfo{year}{2017}\natexlab{}.
\newblock \showarticletitle{$\{$CherryPick$\}$: Adaptively unearthing the best
  cloud configurations for big data analytics}. In \bibinfo{booktitle}{{\em
  14th USENIX Symposium on Networked Systems Design and Implementation (NSDI
  17)}}. \bibinfo{pages}{469--482}.
\newblock


\bibitem[\protect\citeauthoryear{Amazon EC2 Spot Instances}{Amazon EC2 Spot
  Instances}{2023}]%
        {amazon-spot2023}
Amazon EC2 Spot Instances \bibinfo{year}{2023}\natexlab{}.
\newblock \bibinfo{title}{Amazon EC2 Spot Instances}.
\newblock \bibinfo{howpublished}{\url{https://aws.amazon.com/ec2/spot}}.
  (\bibinfo{year}{2023}).
\newblock
\newblock
\shownote{Accessed: 2023-10-21.}


\bibitem[\protect\citeauthoryear{Amazon EC2 Spot Instances Pricing}{Amazon EC2
  Spot Instances Pricing}{2023}]%
        {amazon-spot-pricing2023}
Amazon EC2 Spot Instances Pricing \bibinfo{year}{2023}\natexlab{}.
\newblock \bibinfo{title}{Amazon EC2 Spot Instances Pricing}.
\newblock
  \bibinfo{howpublished}{\url{https://aws.amazon.com/ec2/spot/pricing/}}.
  (\bibinfo{year}{2023}).
\newblock
\newblock
\shownote{Accessed: 2023-10-21.}


\bibitem[\protect\citeauthoryear{Amazon EKS}{Amazon EKS}{2023}]%
        {AmazonEKS}
Amazon EKS \bibinfo{year}{2023}\natexlab{}.
\newblock \bibinfo{title}{Amazon EKS}.
\newblock \bibinfo{howpublished}{\url{https://aws.amazon.com/eks/}}.
  (\bibinfo{year}{2023}).
\newblock
\newblock
\shownote{Accessed: 2023-10-21.}


\bibitem[\protect\citeauthoryear{Amazon RDS}{Amazon RDS}{2023}]%
        {RDS}
Amazon RDS \bibinfo{year}{2023}\natexlab{}.
\newblock \bibinfo{title}{Amazon RDS}.
\newblock \bibinfo{howpublished}{\url{https://aws.amazon.com/rds/}}.
  (\bibinfo{year}{2023}).
\newblock
\newblock
\shownote{Accessed: 2023-10-21.}


\bibitem[\protect\citeauthoryear{Antonopoulos, Budovski, Diaconu,
  Hernandez~Saenz, Hu, Kodavalla, Kossmann, Lingam, Minhas, Prakash,
  et~al\mbox{.}}{Antonopoulos et~al\mbox{.}}{2019}]%
        {antonopoulos2019socrates}
\bibfield{author}{\bibinfo{person}{Panagiotis Antonopoulos},
  \bibinfo{person}{Alex Budovski}, \bibinfo{person}{Cristian Diaconu},
  \bibinfo{person}{Alejandro Hernandez~Saenz}, \bibinfo{person}{Jack Hu},
  \bibinfo{person}{Hanuma Kodavalla}, \bibinfo{person}{Donald Kossmann},
  \bibinfo{person}{Sandeep Lingam}, \bibinfo{person}{Umar~Farooq Minhas},
  \bibinfo{person}{Naveen Prakash}, {et~al\mbox{.}}}
  \bibinfo{year}{2019}\natexlab{}.
\newblock \showarticletitle{Socrates: The new sql server in the cloud}. In
  \bibinfo{booktitle}{{\em Proceedings of the 2019 International Conference on
  Management of Data}}. \bibinfo{pages}{1743--1756}.
\newblock


\bibitem[\protect\citeauthoryear{AWS Lambda}{AWS Lambda}{2023}]%
        {AWSLambda}
AWS Lambda \bibinfo{year}{2023}\natexlab{}.
\newblock \bibinfo{title}{AWS Lambda}.
\newblock \bibinfo{howpublished}{\url{https://aws.amazon.com/lambda/}}.
  (\bibinfo{year}{2023}).
\newblock
\newblock
\shownote{Accessed: 2023-10-21.}


\bibitem[\protect\citeauthoryear{Azure Functions}{Azure Functions}{2023}]%
        {AzureFunctions}
Azure Functions \bibinfo{year}{2023}\natexlab{}.
\newblock \bibinfo{title}{Azure Functions}.
\newblock
  \bibinfo{howpublished}{\url{https://azure.microsoft.com/en-us/products/functions}}.
    (\bibinfo{year}{2023}).
\newblock
\newblock
\shownote{Accessed: 2023-10-21.}


\bibitem[\protect\citeauthoryear{Azure Kubernetes Service}{Azure Kubernetes
  Service}{2023}]%
        {AKS}
Azure Kubernetes Service \bibinfo{year}{2023}\natexlab{}.
\newblock \bibinfo{title}{Azure Kubernetes Service}.
\newblock
  \bibinfo{howpublished}{\url{https://azure.microsoft.com/en-us/products/kubernetes-service}}.
    (\bibinfo{year}{2023}).
\newblock
\newblock
\shownote{Accessed: 2023-10-21.}


\bibitem[\protect\citeauthoryear{Bourtoule, Chandrasekaran, Choquette-Choo,
  Jia, Travers, Zhang, Lie, and Papernot}{Bourtoule et~al\mbox{.}}{2021}]%
        {bourtoule2021machine}
\bibfield{author}{\bibinfo{person}{Lucas Bourtoule}, \bibinfo{person}{Varun
  Chandrasekaran}, \bibinfo{person}{Christopher~A Choquette-Choo},
  \bibinfo{person}{Hengrui Jia}, \bibinfo{person}{Adelin Travers},
  \bibinfo{person}{Baiwu Zhang}, \bibinfo{person}{David Lie}, {and}
  \bibinfo{person}{Nicolas Papernot}.} \bibinfo{year}{2021}\natexlab{}.
\newblock \showarticletitle{Machine unlearning}. In \bibinfo{booktitle}{{\em
  2021 IEEE Symposium on Security and Privacy (SP)}}. IEEE,
  \bibinfo{pages}{141--159}.
\newblock


\bibitem[\protect\citeauthoryear{Bradbury, Frostig, Hawkins, Johnson, Leary,
  Maclaurin, Necula, Paszke, Vander{P}las, Wanderman-{M}ilne, and
  Zhang}{Bradbury et~al\mbox{.}}{2018}]%
        {jax2018github}
\bibfield{author}{\bibinfo{person}{James Bradbury}, \bibinfo{person}{Roy
  Frostig}, \bibinfo{person}{Peter Hawkins}, \bibinfo{person}{Matthew~James
  Johnson}, \bibinfo{person}{Chris Leary}, \bibinfo{person}{Dougal Maclaurin},
  \bibinfo{person}{George Necula}, \bibinfo{person}{Adam Paszke},
  \bibinfo{person}{Jake Vander{P}las}, \bibinfo{person}{Skye
  Wanderman-{M}ilne}, {and} \bibinfo{person}{Qiao Zhang}.}
  \bibinfo{year}{2018}\natexlab{}.
\newblock \bibinfo{title}{{JAX}: composable transformations of
  {P}ython+{N}um{P}y programs}.
\newblock   (\bibinfo{year}{2018}).
\newblock
\showURL{%
\url{http://github.com/google/jax}}


\bibitem[\protect\citeauthoryear{Carbone, Katsifodimos, Ewen, Markl, Haridi,
  and Tzoumas}{Carbone et~al\mbox{.}}{2015}]%
        {carbone2015apache}
\bibfield{author}{\bibinfo{person}{Paris Carbone}, \bibinfo{person}{Asterios
  Katsifodimos}, \bibinfo{person}{Stephan Ewen}, \bibinfo{person}{Volker
  Markl}, \bibinfo{person}{Seif Haridi}, {and} \bibinfo{person}{Kostas
  Tzoumas}.} \bibinfo{year}{2015}\natexlab{}.
\newblock \showarticletitle{Apache flink: Stream and batch processing in a
  single engine}.
\newblock \bibinfo{journal}{{\em The Bulletin of the Technical Committee on
  Data Engineering\/}} \bibinfo{volume}{38}, \bibinfo{number}{4}
  (\bibinfo{year}{2015}).
\newblock


\bibitem[\protect\citeauthoryear{Chaudhuri, Dayal, and Narasayya}{Chaudhuri
  et~al\mbox{.}}{2011}]%
        {chaudhuri2011overview}
\bibfield{author}{\bibinfo{person}{Surajit Chaudhuri},
  \bibinfo{person}{Umeshwar Dayal}, {and} \bibinfo{person}{Vivek Narasayya}.}
  \bibinfo{year}{2011}\natexlab{}.
\newblock \showarticletitle{An overview of business intelligence technology}.
\newblock \bibinfo{journal}{{\it Commun. ACM}} \bibinfo{volume}{54},
  \bibinfo{number}{8} (\bibinfo{year}{2011}), \bibinfo{pages}{88--98}.
\newblock


\bibitem[\protect\citeauthoryear{Chen, Borgeaud, Irving, Lespiau, Sifre, and
  Jumper}{Chen et~al\mbox{.}}{2023}]%
        {chen2023accelerating}
\bibfield{author}{\bibinfo{person}{Charlie Chen}, \bibinfo{person}{Sebastian
  Borgeaud}, \bibinfo{person}{Geoffrey Irving}, \bibinfo{person}{Jean-Baptiste
  Lespiau}, \bibinfo{person}{Laurent Sifre}, {and} \bibinfo{person}{John
  Jumper}.} \bibinfo{year}{2023}\natexlab{}.
\newblock \showarticletitle{Accelerating large language model decoding with
  speculative sampling}.
\newblock \bibinfo{journal}{{\em arXiv preprint arXiv:2302.01318\/}}
  (\bibinfo{year}{2023}).
\newblock


\bibitem[\protect\citeauthoryear{Chen, Ye, Wu, Zhuo, Ceze, and
  Krishnamurthy}{Chen et~al\mbox{.}}{[n. d.]}]%
        {punica}
\bibfield{author}{\bibinfo{person}{Lequn Chen}, \bibinfo{person}{Zihao Ye},
  \bibinfo{person}{Yongji Wu}, \bibinfo{person}{Danyang Zhuo},
  \bibinfo{person}{Luis~Ceze Ceze}, {and} \bibinfo{person}{Arvind
  Krishnamurthy}.} \bibinfo{year}{[n. d.]}\natexlab{}.
\newblock \showarticletitle{Punica: Multi-Tenant LoRA Serving}.
\newblock \bibinfo{journal}{{\em arXiv preprint arXiv:2310.18547\/}}.
\newblock


\bibitem[\protect\citeauthoryear{Chien}{Chien}{2021}]%
        {DBLP:journals/cacm/Chien21a}
\bibfield{author}{\bibinfo{person}{Andrew~A. Chien}.}
  \bibinfo{year}{2021}\natexlab{}.
\newblock \showarticletitle{Driving the Cloud to True Zero Carbon}.
\newblock \bibinfo{journal}{{\it Commun. {ACM}}} \bibinfo{volume}{64},
  \bibinfo{number}{2} (\bibinfo{year}{2021}), \bibinfo{pages}{5}.
\newblock


\bibitem[\protect\citeauthoryear{Dageville, Cruanes, Zukowski, Antonov, Avanes,
  Bock, Claybaugh, Engovatov, Hentschel, Huang, et~al\mbox{.}}{Dageville
  et~al\mbox{.}}{2016}]%
        {dageville2016snowflake}
\bibfield{author}{\bibinfo{person}{Benoit Dageville}, \bibinfo{person}{Thierry
  Cruanes}, \bibinfo{person}{Marcin Zukowski}, \bibinfo{person}{Vadim Antonov},
  \bibinfo{person}{Artin Avanes}, \bibinfo{person}{Jon Bock},
  \bibinfo{person}{Jonathan Claybaugh}, \bibinfo{person}{Daniel Engovatov},
  \bibinfo{person}{Martin Hentschel}, \bibinfo{person}{Jiansheng Huang},
  {et~al\mbox{.}}} \bibinfo{year}{2016}\natexlab{}.
\newblock \showarticletitle{The snowflake elastic data warehouse}. In
  \bibinfo{booktitle}{{\em Proceedings of the 2016 International Conference on
  Management of Data}}. \bibinfo{pages}{215--226}.
\newblock


\bibitem[\protect\citeauthoryear{Dao}{Dao}{2023}]%
        {dao2023flashattention}
\bibfield{author}{\bibinfo{person}{Tri Dao}.} \bibinfo{year}{2023}\natexlab{}.
\newblock \showarticletitle{Flashattention-2: Faster attention with better
  parallelism and work partitioning}.
\newblock \bibinfo{journal}{{\em arXiv preprint arXiv:2307.08691\/}}
  (\bibinfo{year}{2023}).
\newblock


\bibitem[\protect\citeauthoryear{Dao, Fu, Ermon, Rudra, and R{\'e}}{Dao
  et~al\mbox{.}}{2022}]%
        {dao2022flashattention}
\bibfield{author}{\bibinfo{person}{Tri Dao}, \bibinfo{person}{Dan Fu},
  \bibinfo{person}{Stefano Ermon}, \bibinfo{person}{Atri Rudra}, {and}
  \bibinfo{person}{Christopher R{\'e}}.} \bibinfo{year}{2022}\natexlab{}.
\newblock \showarticletitle{Flashattention: Fast and memory-efficient exact
  attention with io-awareness}.
\newblock \bibinfo{journal}{{\em Advances in Neural Information Processing
  Systems\/}}  \bibinfo{volume}{35} (\bibinfo{year}{2022}),
  \bibinfo{pages}{16344--16359}.
\newblock


\bibitem[\protect\citeauthoryear{Databricks}{Databricks}{2023}]%
        {Databricks}
Databricks \bibinfo{year}{2023}\natexlab{}.
\newblock \bibinfo{title}{Databricks}.
\newblock \bibinfo{howpublished}{\url{https://www.databricks.com/}}.
  (\bibinfo{year}{2023}).
\newblock
\newblock
\shownote{Accessed: 2023-10-21.}


\bibitem[\protect\citeauthoryear{DeepSpeed-MII}{DeepSpeed-MII}{2023}]%
        {deepspeedmii}
DeepSpeed-MII \bibinfo{year}{2023}\natexlab{}.
\newblock \bibinfo{title}{DeepSpeed-MII}.
\newblock
  \bibinfo{howpublished}{\url{https://github.com/microsoft/DeepSpeed-MII}}.
  (\bibinfo{year}{2023}).
\newblock
\newblock
\shownote{Accessed: 2023-10-21.}


\bibitem[\protect\citeauthoryear{Fedus, Zoph, and Shazeer}{Fedus
  et~al\mbox{.}}{2022}]%
        {fedus2022switch}
\bibfield{author}{\bibinfo{person}{William Fedus}, \bibinfo{person}{Barret
  Zoph}, {and} \bibinfo{person}{Noam Shazeer}.}
  \bibinfo{year}{2022}\natexlab{}.
\newblock \showarticletitle{Switch transformers: Scaling to trillion parameter
  models with simple and efficient sparsity}.
\newblock \bibinfo{journal}{{\em The Journal of Machine Learning Research\/}}
  \bibinfo{volume}{23}, \bibinfo{number}{1} (\bibinfo{year}{2022}),
  \bibinfo{pages}{5232--5270}.
\newblock


\bibitem[\protect\citeauthoryear{Gan, Zhang, Cheng, Shetty, Rathi, Katarki,
  Bruno, Hu, Ritchken, Jackson, et~al\mbox{.}}{Gan et~al\mbox{.}}{2019}]%
        {gan2019open}
\bibfield{author}{\bibinfo{person}{Yu Gan}, \bibinfo{person}{Yanqi Zhang},
  \bibinfo{person}{Dailun Cheng}, \bibinfo{person}{Ankitha Shetty},
  \bibinfo{person}{Priyal Rathi}, \bibinfo{person}{Nayan Katarki},
  \bibinfo{person}{Ariana Bruno}, \bibinfo{person}{Justin Hu},
  \bibinfo{person}{Brian Ritchken}, \bibinfo{person}{Brendon Jackson},
  {et~al\mbox{.}}} \bibinfo{year}{2019}\natexlab{}.
\newblock \showarticletitle{An open-source benchmark suite for microservices
  and their hardware-software implications for cloud \& edge systems}. In
  \bibinfo{booktitle}{{\em Proceedings of the Twenty-Fourth International
  Conference on Architectural Support for Programming Languages and Operating
  Systems}}. \bibinfo{pages}{3--18}.
\newblock


\bibitem[\protect\citeauthoryear{Ghodsi, Zaharia, Hindman, Konwinski, Shenker,
  and Stoica}{Ghodsi et~al\mbox{.}}{2011}]%
        {ghodsi2011dominant}
\bibfield{author}{\bibinfo{person}{Ali Ghodsi}, \bibinfo{person}{Matei
  Zaharia}, \bibinfo{person}{Benjamin Hindman}, \bibinfo{person}{Andy
  Konwinski}, \bibinfo{person}{Scott Shenker}, {and} \bibinfo{person}{Ion
  Stoica}.} \bibinfo{year}{2011}\natexlab{}.
\newblock \showarticletitle{Dominant resource fairness: Fair allocation of
  multiple resource types}. In \bibinfo{booktitle}{{\em 8th USENIX symposium on
  networked systems design and implementation (NSDI 11)}}.
\newblock


\bibitem[\protect\citeauthoryear{Goldstein and Larson}{Goldstein and
  Larson}{2001}]%
        {goldstein2001optimizing}
\bibfield{author}{\bibinfo{person}{Jonathan Goldstein} {and}
  \bibinfo{person}{Per-{\AA}ke Larson}.} \bibinfo{year}{2001}\natexlab{}.
\newblock \showarticletitle{Optimizing queries using materialized views: a
  practical, scalable solution}.
\newblock \bibinfo{journal}{{\em ACM SIGMOD Record\/}} \bibinfo{volume}{30},
  \bibinfo{number}{2} (\bibinfo{year}{2001}), \bibinfo{pages}{331--342}.
\newblock


\bibitem[\protect\citeauthoryear{Google Cloud Functions}{Google Cloud
  Functions}{2023}]%
        {GoogleCloudFunctions}
Google Cloud Functions \bibinfo{year}{2023}\natexlab{}.
\newblock \bibinfo{title}{Google Cloud Functions}.
\newblock
  \bibinfo{howpublished}{\url{https://cloud.google.com/functions?hl=en}}.
  (\bibinfo{year}{2023}).
\newblock
\newblock
\shownote{Accessed: 2023-10-21.}


\bibitem[\protect\citeauthoryear{Google Kubernetes Engine}{Google Kubernetes
  Engine}{2023}]%
        {GKE}
Google Kubernetes Engine \bibinfo{year}{2023}\natexlab{}.
\newblock \bibinfo{title}{Google Kubernetes Engine}.
\newblock
  \bibinfo{howpublished}{\url{https://cloud.google.com/kubernetes-engine?hl=en}}.
    (\bibinfo{year}{2023}).
\newblock
\newblock
\shownote{Accessed: 2023-10-21.}


\bibitem[\protect\citeauthoryear{Gujarati, Karimi, Alzayat, Hao, Kaufmann,
  Vigfusson, and Mace}{Gujarati et~al\mbox{.}}{2020}]%
        {gujarati2020serving}
\bibfield{author}{\bibinfo{person}{Arpan Gujarati}, \bibinfo{person}{Reza
  Karimi}, \bibinfo{person}{Safya Alzayat}, \bibinfo{person}{Wei Hao},
  \bibinfo{person}{Antoine Kaufmann}, \bibinfo{person}{Ymir Vigfusson}, {and}
  \bibinfo{person}{Jonathan Mace}.} \bibinfo{year}{2020}\natexlab{}.
\newblock \showarticletitle{Serving $\{$DNNs$\}$ like clockwork: Performance
  predictability from the bottom up}. In \bibinfo{booktitle}{{\em 14th USENIX
  Symposium on Operating Systems Design and Implementation (OSDI 20)}}.
  \bibinfo{pages}{443--462}.
\newblock


\bibitem[\protect\citeauthoryear{Gupta, Agarwal, Tan, Kulesza, Pathak, Stefani,
  and Srinivasan}{Gupta et~al\mbox{.}}{2015}]%
        {gupta2015amazon}
\bibfield{author}{\bibinfo{person}{Anurag Gupta}, \bibinfo{person}{Deepak
  Agarwal}, \bibinfo{person}{Derek Tan}, \bibinfo{person}{Jakub Kulesza},
  \bibinfo{person}{Rahul Pathak}, \bibinfo{person}{Stefano Stefani}, {and}
  \bibinfo{person}{Vidhya Srinivasan}.} \bibinfo{year}{2015}\natexlab{}.
\newblock \showarticletitle{Amazon redshift and the case for simpler data
  warehouses}. In \bibinfo{booktitle}{{\em Proceedings of the 2015 ACM SIGMOD
  international conference on management of data}}.
  \bibinfo{pages}{1917--1923}.
\newblock


\bibitem[\protect\citeauthoryear{Hestness, Narang, Ardalani, Diamos, Jun,
  Kianinejad, Patwary, Yang, and Zhou}{Hestness et~al\mbox{.}}{2017}]%
        {hestness2017deep}
\bibfield{author}{\bibinfo{person}{Joel Hestness}, \bibinfo{person}{Sharan
  Narang}, \bibinfo{person}{Newsha Ardalani}, \bibinfo{person}{Gregory Diamos},
  \bibinfo{person}{Heewoo Jun}, \bibinfo{person}{Hassan Kianinejad},
  \bibinfo{person}{Md~Mostofa~Ali Patwary}, \bibinfo{person}{Yang Yang}, {and}
  \bibinfo{person}{Yanqi Zhou}.} \bibinfo{year}{2017}\natexlab{}.
\newblock \showarticletitle{Deep learning scaling is predictable, empirically}.
\newblock \bibinfo{journal}{{\em arXiv preprint arXiv:1712.00409\/}}
  (\bibinfo{year}{2017}).
\newblock


\bibitem[\protect\citeauthoryear{Hindman, Konwinski, Zaharia, Ghodsi, Joseph,
  Katz, Shenker, and Stoica}{Hindman et~al\mbox{.}}{2011}]%
        {hindman2011mesos}
\bibfield{author}{\bibinfo{person}{Benjamin Hindman}, \bibinfo{person}{Andy
  Konwinski}, \bibinfo{person}{Matei Zaharia}, \bibinfo{person}{Ali Ghodsi},
  \bibinfo{person}{Anthony~D Joseph}, \bibinfo{person}{Randy Katz},
  \bibinfo{person}{Scott Shenker}, {and} \bibinfo{person}{Ion Stoica}.}
  \bibinfo{year}{2011}\natexlab{}.
\newblock \showarticletitle{Mesos: A platform for $\{$Fine-Grained$\}$ resource
  sharing in the data center}. In \bibinfo{booktitle}{{\em 8th USENIX Symposium
  on Networked Systems Design and Implementation (NSDI 11)}}.
\newblock


\bibitem[\protect\citeauthoryear{Hong, Zheng, Chen, Cheng, Zhang, Wang, Yau,
  Lin, Zhou, Ran, Xiao, and Wu}{Hong et~al\mbox{.}}{2023}]%
        {Hong2023MetaGPTMP}
\bibfield{author}{\bibinfo{person}{Sirui Hong}, \bibinfo{person}{Xiawu Zheng},
  \bibinfo{person}{Jonathan~P. Chen}, \bibinfo{person}{Yuheng Cheng},
  \bibinfo{person}{Ceyao Zhang}, \bibinfo{person}{Zili Wang},
  \bibinfo{person}{Steven Ka~Shing Yau}, \bibinfo{person}{Zi~Hen Lin},
  \bibinfo{person}{Liyang Zhou}, \bibinfo{person}{Chenyu Ran},
  \bibinfo{person}{Lingfeng Xiao}, {and} \bibinfo{person}{Chenglin Wu}.}
  \bibinfo{year}{2023}\natexlab{}.
\newblock \showarticletitle{MetaGPT: Meta Programming for Multi-Agent
  Collaborative Framework}.
\newblock \bibinfo{journal}{{\em ArXiv\/}}  \bibinfo{volume}{abs/2308.00352}
  (\bibinfo{year}{2023}).
\newblock
\showURL{%
\url{https://api.semanticscholar.org/CorpusID:260351380}}


\bibitem[\protect\citeauthoryear{Hu, Shen, Wallis, Allen-Zhu, Li, Wang, Wang,
  and Chen}{Hu et~al\mbox{.}}{2021}]%
        {hu2021lora}
\bibfield{author}{\bibinfo{person}{Edward~J Hu}, \bibinfo{person}{Yelong Shen},
  \bibinfo{person}{Phillip Wallis}, \bibinfo{person}{Zeyuan Allen-Zhu},
  \bibinfo{person}{Yuanzhi Li}, \bibinfo{person}{Shean Wang},
  \bibinfo{person}{Lu Wang}, {and} \bibinfo{person}{Weizhu Chen}.}
  \bibinfo{year}{2021}\natexlab{}.
\newblock \showarticletitle{Lora: Low-rank adaptation of large language
  models}.
\newblock \bibinfo{journal}{{\em arXiv preprint arXiv:2106.09685\/}}
  (\bibinfo{year}{2021}).
\newblock


\bibitem[\protect\citeauthoryear{Huang, Cheng, Bapna, Firat, Chen, Chen, Lee,
  Ngiam, Le, Wu, et~al\mbox{.}}{Huang et~al\mbox{.}}{2019}]%
        {huang2019gpipe}
\bibfield{author}{\bibinfo{person}{Yanping Huang}, \bibinfo{person}{Youlong
  Cheng}, \bibinfo{person}{Ankur Bapna}, \bibinfo{person}{Orhan Firat},
  \bibinfo{person}{Dehao Chen}, \bibinfo{person}{Mia Chen},
  \bibinfo{person}{HyoukJoong Lee}, \bibinfo{person}{Jiquan Ngiam},
  \bibinfo{person}{Quoc~V Le}, \bibinfo{person}{Yonghui Wu}, {et~al\mbox{.}}}
  \bibinfo{year}{2019}\natexlab{}.
\newblock \showarticletitle{Gpipe: Efficient training of giant neural networks
  using pipeline parallelism}.
\newblock \bibinfo{journal}{{\em Advances in neural information processing
  systems\/}}  \bibinfo{volume}{32} (\bibinfo{year}{2019}).
\newblock


\bibitem[\protect\citeauthoryear{Jonas, Pu, Venkataraman, Stoica, and
  Recht}{Jonas et~al\mbox{.}}{2017}]%
        {jonas2017occupy}
\bibfield{author}{\bibinfo{person}{Eric Jonas}, \bibinfo{person}{Qifan Pu},
  \bibinfo{person}{Shivaram Venkataraman}, \bibinfo{person}{Ion Stoica}, {and}
  \bibinfo{person}{Benjamin Recht}.} \bibinfo{year}{2017}\natexlab{}.
\newblock \showarticletitle{Occupy the cloud: Distributed computing for the
  99\%}. In \bibinfo{booktitle}{{\em Proceedings of the 2017 symposium on cloud
  computing}}. \bibinfo{pages}{445--451}.
\newblock


\bibitem[\protect\citeauthoryear{Jonas, Schleier-Smith, Sreekanti, Tsai,
  Khandelwal, Pu, Shankar, Carreira, Krauth, Yadwadkar, et~al\mbox{.}}{Jonas
  et~al\mbox{.}}{2019}]%
        {jonas2019cloud}
\bibfield{author}{\bibinfo{person}{Eric Jonas}, \bibinfo{person}{Johann
  Schleier-Smith}, \bibinfo{person}{Vikram Sreekanti},
  \bibinfo{person}{Chia-Che Tsai}, \bibinfo{person}{Anurag Khandelwal},
  \bibinfo{person}{Qifan Pu}, \bibinfo{person}{Vaishaal Shankar},
  \bibinfo{person}{Joao Carreira}, \bibinfo{person}{Karl Krauth},
  \bibinfo{person}{Neeraja Yadwadkar}, {et~al\mbox{.}}}
  \bibinfo{year}{2019}\natexlab{}.
\newblock \showarticletitle{Cloud programming simplified: A berkeley view on
  serverless computing}.
\newblock \bibinfo{journal}{{\em arXiv preprint arXiv:1902.03383\/}}
  (\bibinfo{year}{2019}).
\newblock


\bibitem[\protect\citeauthoryear{Kubernetes}{Kubernetes}{2023}]%
        {kubernetes}
Kubernetes \bibinfo{year}{2023}\natexlab{}.
\newblock \bibinfo{title}{Kubernetes}.
\newblock \bibinfo{howpublished}{\url{https://kubernetes.io/}}.
  (\bibinfo{year}{2023}).
\newblock
\newblock
\shownote{Accessed: 2023-10-21.}


\bibitem[\protect\citeauthoryear{Kwon, Li, Zhuang, Sheng, Zheng, Yu, Gonzalez,
  Zhang, and Stoica}{Kwon et~al\mbox{.}}{2023}]%
        {kwon2023efficient}
\bibfield{author}{\bibinfo{person}{Woosuk Kwon}, \bibinfo{person}{Zhuohan Li},
  \bibinfo{person}{Siyuan Zhuang}, \bibinfo{person}{Ying Sheng},
  \bibinfo{person}{Lianmin Zheng}, \bibinfo{person}{Cody~Hao Yu},
  \bibinfo{person}{Joseph~E Gonzalez}, \bibinfo{person}{Hao Zhang}, {and}
  \bibinfo{person}{Ion Stoica}.} \bibinfo{year}{2023}\natexlab{}.
\newblock \showarticletitle{Efficient memory management for large language
  model serving with pagedattention}.
\newblock \bibinfo{journal}{{\em arXiv preprint arXiv:2309.06180\/}}
  (\bibinfo{year}{2023}).
\newblock


\bibitem[\protect\citeauthoryear{Leviathan, Kalman, and Matias}{Leviathan
  et~al\mbox{.}}{2023}]%
        {leviathan2023fast}
\bibfield{author}{\bibinfo{person}{Yaniv Leviathan}, \bibinfo{person}{Matan
  Kalman}, {and} \bibinfo{person}{Yossi Matias}.}
  \bibinfo{year}{2023}\natexlab{}.
\newblock \showarticletitle{Fast inference from transformers via speculative
  decoding}. In \bibinfo{booktitle}{{\em International Conference on Machine
  Learning}}. PMLR, \bibinfo{pages}{19274--19286}.
\newblock


\bibitem[\protect\citeauthoryear{Lewis, Perez, Piktus, Petroni, Karpukhin,
  Goyal, K{\"u}ttler, Lewis, Yih, Rockt{\"a}schel, et~al\mbox{.}}{Lewis
  et~al\mbox{.}}{2020}]%
        {lewis2020retrieval}
\bibfield{author}{\bibinfo{person}{Patrick Lewis}, \bibinfo{person}{Ethan
  Perez}, \bibinfo{person}{Aleksandra Piktus}, \bibinfo{person}{Fabio Petroni},
  \bibinfo{person}{Vladimir Karpukhin}, \bibinfo{person}{Naman Goyal},
  \bibinfo{person}{Heinrich K{\"u}ttler}, \bibinfo{person}{Mike Lewis},
  \bibinfo{person}{Wen-tau Yih}, \bibinfo{person}{Tim Rockt{\"a}schel},
  {et~al\mbox{.}}} \bibinfo{year}{2020}\natexlab{}.
\newblock \showarticletitle{Retrieval-augmented generation for
  knowledge-intensive nlp tasks}.
\newblock \bibinfo{journal}{{\em Advances in Neural Information Processing
  Systems\/}}  \bibinfo{volume}{33} (\bibinfo{year}{2020}),
  \bibinfo{pages}{9459--9474}.
\newblock


\bibitem[\protect\citeauthoryear{Li, Lu, and Kandula}{Li et~al\mbox{.}}{2022}]%
        {li2022warper}
\bibfield{author}{\bibinfo{person}{Beibin Li}, \bibinfo{person}{Yao Lu}, {and}
  \bibinfo{person}{Srikanth Kandula}.} \bibinfo{year}{2022}\natexlab{}.
\newblock \showarticletitle{Warper: Efficiently adapting learned cardinality
  estimators to data and workload drifts}. In \bibinfo{booktitle}{{\em
  Proceedings of the 2022 International Conference on Management of Data}}.
  \bibinfo{pages}{1920--1933}.
\newblock


\bibitem[\protect\citeauthoryear{Liu, Yu, Zhang, Xu, Lei, Lai, Gu, Ding, Men,
  Yang, Zhang, Deng, Zeng, Du, Zhang, Shen, Zhang, Su, Sun, Huang, Dong, and
  Tang}{Liu et~al\mbox{.}}{2023}]%
        {liu2023agentbench}
\bibfield{author}{\bibinfo{person}{Xiao Liu}, \bibinfo{person}{Hao Yu},
  \bibinfo{person}{Hanchen Zhang}, \bibinfo{person}{Yifan Xu},
  \bibinfo{person}{Xuanyu Lei}, \bibinfo{person}{Hanyu Lai},
  \bibinfo{person}{Yu Gu}, \bibinfo{person}{Hangliang Ding},
  \bibinfo{person}{Kaiwen Men}, \bibinfo{person}{Kejuan Yang},
  \bibinfo{person}{Shudan Zhang}, \bibinfo{person}{Xiang Deng},
  \bibinfo{person}{Aohan Zeng}, \bibinfo{person}{Zhengxiao Du},
  \bibinfo{person}{Chenhui Zhang}, \bibinfo{person}{Sheng Shen},
  \bibinfo{person}{Tianjun Zhang}, \bibinfo{person}{Yu Su},
  \bibinfo{person}{Huan Sun}, \bibinfo{person}{Minlie Huang},
  \bibinfo{person}{Yuxiao Dong}, {and} \bibinfo{person}{Jie Tang}.}
  \bibinfo{year}{2023}\natexlab{}.
\newblock \showarticletitle{AgentBench: Evaluating LLMs as Agents}.
\newblock \bibinfo{journal}{{\em arXiv preprint arXiv: 2308.03688\/}}
  (\bibinfo{year}{2023}).
\newblock


\bibitem[\protect\citeauthoryear{Ma, Van~Aken, Hefny, Mezerhane, Pavlo, and
  Gordon}{Ma et~al\mbox{.}}{2018}]%
        {ma2018query}
\bibfield{author}{\bibinfo{person}{Lin Ma}, \bibinfo{person}{Dana Van~Aken},
  \bibinfo{person}{Ahmed Hefny}, \bibinfo{person}{Gustavo Mezerhane},
  \bibinfo{person}{Andrew Pavlo}, {and} \bibinfo{person}{Geoffrey~J Gordon}.}
  \bibinfo{year}{2018}\natexlab{}.
\newblock \showarticletitle{Query-based workload forecasting for self-driving
  database management systems}. In \bibinfo{booktitle}{{\em Proceedings of the
  2018 International Conference on Management of Data}}.
  \bibinfo{pages}{631--645}.
\newblock


\bibitem[\protect\citeauthoryear{Mahajan, Balasubramanian, Singhvi,
  Venkataraman, Akella, Phanishayee, and Chawla}{Mahajan et~al\mbox{.}}{2020}]%
        {mahajan2020themis}
\bibfield{author}{\bibinfo{person}{Kshiteej Mahajan}, \bibinfo{person}{Arjun
  Balasubramanian}, \bibinfo{person}{Arjun Singhvi}, \bibinfo{person}{Shivaram
  Venkataraman}, \bibinfo{person}{Aditya Akella}, \bibinfo{person}{Amar
  Phanishayee}, {and} \bibinfo{person}{Shuchi Chawla}.}
  \bibinfo{year}{2020}\natexlab{}.
\newblock \showarticletitle{Themis: Fair and efficient $\{$GPU$\}$ cluster
  scheduling}. In \bibinfo{booktitle}{{\em 17th USENIX Symposium on Networked
  Systems Design and Implementation (NSDI 20)}}. \bibinfo{pages}{289--304}.
\newblock


\bibitem[\protect\citeauthoryear{Malkov and Yashunin}{Malkov and
  Yashunin}{2018}]%
        {malkov2018efficient}
\bibfield{author}{\bibinfo{person}{Yu~A Malkov} {and} \bibinfo{person}{Dmitry~A
  Yashunin}.} \bibinfo{year}{2018}\natexlab{}.
\newblock \showarticletitle{Efficient and robust approximate nearest neighbor
  search using hierarchical navigable small world graphs}.
\newblock \bibinfo{journal}{{\em IEEE transactions on pattern analysis and
  machine intelligence\/}} \bibinfo{volume}{42}, \bibinfo{number}{4}
  (\bibinfo{year}{2018}), \bibinfo{pages}{824--836}.
\newblock


\bibitem[\protect\citeauthoryear{Mohoney, Pacaci, Chowdhury, Mousavi, Ilyas,
  Minhas, Pound, and Rekatsinas}{Mohoney et~al\mbox{.}}{2023}]%
        {mohoney2023high}
\bibfield{author}{\bibinfo{person}{Jason Mohoney}, \bibinfo{person}{Anil
  Pacaci}, \bibinfo{person}{Shihabur~Rahman Chowdhury}, \bibinfo{person}{Ali
  Mousavi}, \bibinfo{person}{Ihab~F Ilyas}, \bibinfo{person}{Umar~Farooq
  Minhas}, \bibinfo{person}{Jeffrey Pound}, {and} \bibinfo{person}{Theodoros
  Rekatsinas}.} \bibinfo{year}{2023}\natexlab{}.
\newblock \showarticletitle{High-Throughput Vector Similarity Search in
  Knowledge Graphs}.
\newblock \bibinfo{journal}{{\em Proceedings of the ACM on Management of
  Data\/}} \bibinfo{volume}{1}, \bibinfo{number}{2} (\bibinfo{year}{2023}),
  \bibinfo{pages}{1--25}.
\newblock


\bibitem[\protect\citeauthoryear{Murray, McSherry, Isaacs, Isard, Barham, and
  Abadi}{Murray et~al\mbox{.}}{2013}]%
        {murray2013naiad}
\bibfield{author}{\bibinfo{person}{Derek~G. Murray}, \bibinfo{person}{Frank
  McSherry}, \bibinfo{person}{Rebecca Isaacs}, \bibinfo{person}{Michael Isard},
  \bibinfo{person}{Paul Barham}, {and} \bibinfo{person}{Mart\'{\i}n Abadi}.}
  \bibinfo{year}{2013}\natexlab{}.
\newblock \showarticletitle{{Naiad: A Timely Dataflow System}}. In
  \bibinfo{booktitle}{{\em ACM Symposium on Operating Systems Principles
  (SOSP)}}.
\newblock


\bibitem[\protect\citeauthoryear{Narayanan, Harlap, Phanishayee, Seshadri,
  Devanur, Ganger, Gibbons, and Zaharia}{Narayanan et~al\mbox{.}}{2019}]%
        {narayanan2019pipedream}
\bibfield{author}{\bibinfo{person}{Deepak Narayanan}, \bibinfo{person}{Aaron
  Harlap}, \bibinfo{person}{Amar Phanishayee}, \bibinfo{person}{Vivek
  Seshadri}, \bibinfo{person}{Nikhil~R Devanur}, \bibinfo{person}{Gregory~R
  Ganger}, \bibinfo{person}{Phillip~B Gibbons}, {and} \bibinfo{person}{Matei
  Zaharia}.} \bibinfo{year}{2019}\natexlab{}.
\newblock \showarticletitle{PipeDream: Generalized pipeline parallelism for DNN
  training}. In \bibinfo{booktitle}{{\em Proceedings of the 27th ACM Symposium
  on Operating Systems Principles}}. \bibinfo{pages}{1--15}.
\newblock


\bibitem[\protect\citeauthoryear{Narayanan, Phanishayee, Shi, Chen, and
  Zaharia}{Narayanan et~al\mbox{.}}{2021a}]%
        {narayanan2021memory}
\bibfield{author}{\bibinfo{person}{Deepak Narayanan}, \bibinfo{person}{Amar
  Phanishayee}, \bibinfo{person}{Kaiyu Shi}, \bibinfo{person}{Xie Chen}, {and}
  \bibinfo{person}{Matei Zaharia}.} \bibinfo{year}{2021}\natexlab{a}.
\newblock \showarticletitle{Memory-efficient pipeline-parallel dnn training}.
  In \bibinfo{booktitle}{{\em International Conference on Machine Learning}}.
  PMLR, \bibinfo{pages}{7937--7947}.
\newblock


\bibitem[\protect\citeauthoryear{Narayanan, Santhanam, Kazhamiaka, Phanishayee,
  and Zaharia}{Narayanan et~al\mbox{.}}{2020}]%
        {narayanan2020heterogeneity}
\bibfield{author}{\bibinfo{person}{Deepak Narayanan}, \bibinfo{person}{Keshav
  Santhanam}, \bibinfo{person}{Fiodar Kazhamiaka}, \bibinfo{person}{Amar
  Phanishayee}, {and} \bibinfo{person}{Matei Zaharia}.}
  \bibinfo{year}{2020}\natexlab{}.
\newblock \showarticletitle{$\{$Heterogeneity-Aware$\}$ cluster scheduling
  policies for deep learning workloads}. In \bibinfo{booktitle}{{\em 14th
  USENIX Symposium on Operating Systems Design and Implementation (OSDI 20)}}.
  \bibinfo{pages}{481--498}.
\newblock


\bibitem[\protect\citeauthoryear{Narayanan, Shoeybi, Casper, LeGresley,
  Patwary, Korthikanti, Vainbrand, Kashinkunti, Bernauer, Catanzaro,
  et~al\mbox{.}}{Narayanan et~al\mbox{.}}{2021b}]%
        {narayanan2021efficient}
\bibfield{author}{\bibinfo{person}{Deepak Narayanan}, \bibinfo{person}{Mohammad
  Shoeybi}, \bibinfo{person}{Jared Casper}, \bibinfo{person}{Patrick
  LeGresley}, \bibinfo{person}{Mostofa Patwary}, \bibinfo{person}{Vijay
  Korthikanti}, \bibinfo{person}{Dmitri Vainbrand}, \bibinfo{person}{Prethvi
  Kashinkunti}, \bibinfo{person}{Julie Bernauer}, \bibinfo{person}{Bryan
  Catanzaro}, {et~al\mbox{.}}} \bibinfo{year}{2021}\natexlab{b}.
\newblock \showarticletitle{Efficient large-scale language model training on
  gpu clusters using megatron-lm}. In \bibinfo{booktitle}{{\em Proceedings of
  the International Conference for High Performance Computing, Networking,
  Storage and Analysis}}. \bibinfo{pages}{1--15}.
\newblock


\bibitem[\protect\citeauthoryear{Olston, Fiedel, Gorovoy, Harmsen, Lao, Li,
  Rajashekhar, Ramesh, and Soyke}{Olston et~al\mbox{.}}{2017}]%
        {olston2017tensorflow}
\bibfield{author}{\bibinfo{person}{Christopher Olston}, \bibinfo{person}{Noah
  Fiedel}, \bibinfo{person}{Kiril Gorovoy}, \bibinfo{person}{Jeremiah Harmsen},
  \bibinfo{person}{Li Lao}, \bibinfo{person}{Fangwei Li}, \bibinfo{person}{Vinu
  Rajashekhar}, \bibinfo{person}{Sukriti Ramesh}, {and} \bibinfo{person}{Jordan
  Soyke}.} \bibinfo{year}{2017}\natexlab{}.
\newblock \showarticletitle{Tensorflow-serving: Flexible, high-performance ml
  serving}.
\newblock \bibinfo{journal}{{\em arXiv preprint arXiv:1712.06139\/}}
  (\bibinfo{year}{2017}).
\newblock


\bibitem[\protect\citeauthoryear{Ouyang, Wu, Jiang, Almeida, Wainwright,
  Mishkin, Zhang, Agarwal, Slama, Ray, et~al\mbox{.}}{Ouyang
  et~al\mbox{.}}{2022}]%
        {ouyang2022training}
\bibfield{author}{\bibinfo{person}{Long Ouyang}, \bibinfo{person}{Jeffrey Wu},
  \bibinfo{person}{Xu Jiang}, \bibinfo{person}{Diogo Almeida},
  \bibinfo{person}{Carroll Wainwright}, \bibinfo{person}{Pamela Mishkin},
  \bibinfo{person}{Chong Zhang}, \bibinfo{person}{Sandhini Agarwal},
  \bibinfo{person}{Katarina Slama}, \bibinfo{person}{Alex Ray},
  {et~al\mbox{.}}} \bibinfo{year}{2022}\natexlab{}.
\newblock \showarticletitle{Training language models to follow instructions
  with human feedback}.
\newblock \bibinfo{journal}{{\em Advances in Neural Information Processing
  Systems\/}}  \bibinfo{volume}{35} (\bibinfo{year}{2022}),
  \bibinfo{pages}{27730--27744}.
\newblock


\bibitem[\protect\citeauthoryear{Pan, Wang, and Li}{Pan et~al\mbox{.}}{2023}]%
        {VecDBSurvey23}
\bibfield{author}{\bibinfo{person}{James~Jie Pan}, \bibinfo{person}{Jianguo
  Wang}, {and} \bibinfo{person}{Guoliang Li}.} \bibinfo{year}{2023}\natexlab{}.
\newblock \showarticletitle{{Survey of Vector Database Management Systems}}.
\newblock \bibinfo{journal}{{\em CoRR\/}}  \bibinfo{volume}{abs/2305.01087}
  (\bibinfo{year}{2023}).
\newblock


\bibitem[\protect\citeauthoryear{Park, O'Brien, Cai, Morris, Liang, and
  Bernstein}{Park et~al\mbox{.}}{2023}]%
        {park2023generative}
\bibfield{author}{\bibinfo{person}{Joon~Sung Park}, \bibinfo{person}{Joseph~C.
  O'Brien}, \bibinfo{person}{Carrie~J. Cai}, \bibinfo{person}{Meredith~Ringel
  Morris}, \bibinfo{person}{Percy Liang}, {and} \bibinfo{person}{Michael~S.
  Bernstein}.} \bibinfo{year}{2023}\natexlab{}.
\newblock \bibinfo{title}{Generative Agents: Interactive Simulacra of Human
  Behavior}.
\newblock   (\bibinfo{year}{2023}).
\newblock
\showeprint[arxiv]{cs.HC/2304.03442}


\bibitem[\protect\citeauthoryear{Paszke, Gross, Massa, Lerer, Bradbury, Chanan,
  Killeen, Lin, Gimelshein, Antiga, et~al\mbox{.}}{Paszke
  et~al\mbox{.}}{2019}]%
        {paszke2019pytorch}
\bibfield{author}{\bibinfo{person}{Adam Paszke}, \bibinfo{person}{Sam Gross},
  \bibinfo{person}{Francisco Massa}, \bibinfo{person}{Adam Lerer},
  \bibinfo{person}{James Bradbury}, \bibinfo{person}{Gregory Chanan},
  \bibinfo{person}{Trevor Killeen}, \bibinfo{person}{Zeming Lin},
  \bibinfo{person}{Natalia Gimelshein}, \bibinfo{person}{Luca Antiga},
  {et~al\mbox{.}}} \bibinfo{year}{2019}\natexlab{}.
\newblock \showarticletitle{Pytorch: An imperative style, high-performance deep
  learning library}.
\newblock \bibinfo{journal}{{\em Advances in Neural Information Processing
  Systems (NeurIPS)\/}}  \bibinfo{volume}{32} (\bibinfo{year}{2019}).
\newblock


\bibitem[\protect\citeauthoryear{Patterson, Gonzalez, Le, Liang, Munguia,
  Rothchild, So, Texier, and Dean}{Patterson et~al\mbox{.}}{2021}]%
        {patterson2021carbon}
\bibfield{author}{\bibinfo{person}{David Patterson}, \bibinfo{person}{Joseph
  Gonzalez}, \bibinfo{person}{Quoc Le}, \bibinfo{person}{Chen Liang},
  \bibinfo{person}{Lluis-Miquel Munguia}, \bibinfo{person}{Daniel Rothchild},
  \bibinfo{person}{David So}, \bibinfo{person}{Maud Texier}, {and}
  \bibinfo{person}{Jeff Dean}.} \bibinfo{year}{2021}\natexlab{}.
\newblock \showarticletitle{Carbon emissions and large neural network
  training}.
\newblock \bibinfo{journal}{{\em arXiv preprint arXiv:2104.10350\/}}
  (\bibinfo{year}{2021}).
\newblock


\bibitem[\protect\citeauthoryear{Peng, Bao, Chen, Wu, and Guo}{Peng
  et~al\mbox{.}}{2018}]%
        {peng2018optimus}
\bibfield{author}{\bibinfo{person}{Yanghua Peng}, \bibinfo{person}{Yixin Bao},
  \bibinfo{person}{Yangrui Chen}, \bibinfo{person}{Chuan Wu}, {and}
  \bibinfo{person}{Chuanxiong Guo}.} \bibinfo{year}{2018}\natexlab{}.
\newblock \showarticletitle{Optimus: an efficient dynamic resource scheduler
  for deep learning clusters}. In \bibinfo{booktitle}{{\em Proceedings of the
  Thirteenth EuroSys Conference}}. \bibinfo{pages}{1--14}.
\newblock


\bibitem[\protect\citeauthoryear{Pope, Douglas, Chowdhery, Devlin, Bradbury,
  Heek, Xiao, Agrawal, and Dean}{Pope et~al\mbox{.}}{2023}]%
        {pope2023efficiently}
\bibfield{author}{\bibinfo{person}{Reiner Pope}, \bibinfo{person}{Sholto
  Douglas}, \bibinfo{person}{Aakanksha Chowdhery}, \bibinfo{person}{Jacob
  Devlin}, \bibinfo{person}{James Bradbury}, \bibinfo{person}{Jonathan Heek},
  \bibinfo{person}{Kefan Xiao}, \bibinfo{person}{Shivani Agrawal}, {and}
  \bibinfo{person}{Jeff Dean}.} \bibinfo{year}{2023}\natexlab{}.
\newblock \showarticletitle{Efficiently scaling transformer inference}.
\newblock \bibinfo{journal}{{\em Proceedings of Machine Learning and
  Systems\/}}  \bibinfo{volume}{5} (\bibinfo{year}{2023}).
\newblock


\bibitem[\protect\citeauthoryear{Rajbhandari, Li, Yao, Zhang, Aminabadi, Awan,
  Rasley, and He}{Rajbhandari et~al\mbox{.}}{2022}]%
        {rajbhandari2022deepspeed}
\bibfield{author}{\bibinfo{person}{Samyam Rajbhandari},
  \bibinfo{person}{Conglong Li}, \bibinfo{person}{Zhewei Yao},
  \bibinfo{person}{Minjia Zhang}, \bibinfo{person}{Reza~Yazdani Aminabadi},
  \bibinfo{person}{Ammar~Ahmad Awan}, \bibinfo{person}{Jeff Rasley}, {and}
  \bibinfo{person}{Yuxiong He}.} \bibinfo{year}{2022}\natexlab{}.
\newblock \showarticletitle{Deepspeed-moe: Advancing mixture-of-experts
  inference and training to power next-generation ai scale}. In
  \bibinfo{booktitle}{{\em International Conference on Machine Learning}}.
  PMLR, \bibinfo{pages}{18332--18346}.
\newblock


\bibitem[\protect\citeauthoryear{Rasley, Rajbhandari, Ruwase, and He}{Rasley
  et~al\mbox{.}}{2020}]%
        {rasley2020deepspeed}
\bibfield{author}{\bibinfo{person}{Jeff Rasley}, \bibinfo{person}{Samyam
  Rajbhandari}, \bibinfo{person}{Olatunji Ruwase}, {and}
  \bibinfo{person}{Yuxiong He}.} \bibinfo{year}{2020}\natexlab{}.
\newblock \showarticletitle{Deepspeed: System optimizations enable training
  deep learning models with over 100 billion parameters}. In
  \bibinfo{booktitle}{{\em Proceedings of the 26th ACM SIGKDD International
  Conference on Knowledge Discovery \& Data Mining}}.
  \bibinfo{pages}{3505--3506}.
\newblock


\bibitem[\protect\citeauthoryear{Shafiei, Khonsari, and Mousavi}{Shafiei
  et~al\mbox{.}}{2022}]%
        {DBLP:journals/csur/ShafieiKM22}
\bibfield{author}{\bibinfo{person}{Hossein Shafiei}, \bibinfo{person}{Ahmad
  Khonsari}, {and} \bibinfo{person}{Payam Mousavi}.}
  \bibinfo{year}{2022}\natexlab{}.
\newblock \showarticletitle{Serverless Computing: {A} Survey of Opportunities,
  Challenges, and Applications}.
\newblock \bibinfo{journal}{{\em {ACM} Computing Survey\/}}
  \bibinfo{volume}{54}, \bibinfo{number}{11s} (\bibinfo{year}{2022}),
  \bibinfo{pages}{239:1--239:32}.
\newblock


\bibitem[\protect\citeauthoryear{Shazeer, Mirhoseini, Maziarz, Davis, Le,
  Hinton, and Dean}{Shazeer et~al\mbox{.}}{2017}]%
        {shazeer2017outrageously}
\bibfield{author}{\bibinfo{person}{Noam Shazeer}, \bibinfo{person}{Azalia
  Mirhoseini}, \bibinfo{person}{Krzysztof Maziarz}, \bibinfo{person}{Andy
  Davis}, \bibinfo{person}{Quoc Le}, \bibinfo{person}{Geoffrey Hinton}, {and}
  \bibinfo{person}{Jeff Dean}.} \bibinfo{year}{2017}\natexlab{}.
\newblock \showarticletitle{Outrageously large neural networks: The
  sparsely-gated mixture-of-experts layer}.
\newblock \bibinfo{journal}{{\em arXiv preprint arXiv:1701.06538\/}}
  (\bibinfo{year}{2017}).
\newblock


\bibitem[\protect\citeauthoryear{Shen, Chen, Jin, Zhao, Kong, Philipose,
  Krishnamurthy, and Sundaram}{Shen et~al\mbox{.}}{2019}]%
        {shen2019nexus}
\bibfield{author}{\bibinfo{person}{Haichen Shen}, \bibinfo{person}{Lequn Chen},
  \bibinfo{person}{Yuchen Jin}, \bibinfo{person}{Liangyu Zhao},
  \bibinfo{person}{Bingyu Kong}, \bibinfo{person}{Matthai Philipose},
  \bibinfo{person}{Arvind Krishnamurthy}, {and} \bibinfo{person}{Ravi
  Sundaram}.} \bibinfo{year}{2019}\natexlab{}.
\newblock \showarticletitle{Nexus: a GPU cluster engine for accelerating
  DNN-based video analysis}. In \bibinfo{booktitle}{{\em ACM Symposium on
  Operating Systems Principles (SOSP)}}.
\newblock


\bibitem[\protect\citeauthoryear{Sheng, Zheng, Yuan, Li, Ryabinin, Chen, Liang,
  Re, Stoica, and Zhang}{Sheng et~al\mbox{.}}{2023}]%
        {sheng2023flexgen}
\bibfield{author}{\bibinfo{person}{Ying Sheng}, \bibinfo{person}{Lianmin
  Zheng}, \bibinfo{person}{Binhang Yuan}, \bibinfo{person}{Zhuohan Li},
  \bibinfo{person}{Max Ryabinin}, \bibinfo{person}{Beidi Chen},
  \bibinfo{person}{Percy Liang}, \bibinfo{person}{Christopher Re},
  \bibinfo{person}{Ion Stoica}, {and} \bibinfo{person}{Ce Zhang}.}
  \bibinfo{year}{2023}\natexlab{}.
\newblock \showarticletitle{FlexGen: High-Throughput Generative Inference of
  Large Language Models with a Single GPU}.
\newblock  (\bibinfo{year}{2023}).
\newblock


\bibitem[\protect\citeauthoryear{Shoeybi, Patwary, Puri, LeGresley, Casper, and
  Catanzaro}{Shoeybi et~al\mbox{.}}{2019}]%
        {shoeybi2019megatron}
\bibfield{author}{\bibinfo{person}{Mohammad Shoeybi}, \bibinfo{person}{Mostofa
  Patwary}, \bibinfo{person}{Raul Puri}, \bibinfo{person}{Patrick LeGresley},
  \bibinfo{person}{Jared Casper}, {and} \bibinfo{person}{Bryan Catanzaro}.}
  \bibinfo{year}{2019}\natexlab{}.
\newblock \showarticletitle{Megatron-lm: Training multi-billion parameter
  language models using model parallelism}.
\newblock \bibinfo{journal}{{\em arXiv preprint arXiv:1909.08053\/}}
  (\bibinfo{year}{2019}).
\newblock


\bibitem[\protect\citeauthoryear{Smith, Patwary, Norick, LeGresley,
  Rajbhandari, Casper, Liu, Prabhumoye, Zerveas, Korthikanti,
  et~al\mbox{.}}{Smith et~al\mbox{.}}{2022}]%
        {smith2022using}
\bibfield{author}{\bibinfo{person}{Shaden Smith}, \bibinfo{person}{Mostofa
  Patwary}, \bibinfo{person}{Brandon Norick}, \bibinfo{person}{Patrick
  LeGresley}, \bibinfo{person}{Samyam Rajbhandari}, \bibinfo{person}{Jared
  Casper}, \bibinfo{person}{Zhun Liu}, \bibinfo{person}{Shrimai Prabhumoye},
  \bibinfo{person}{George Zerveas}, \bibinfo{person}{Vijay Korthikanti},
  {et~al\mbox{.}}} \bibinfo{year}{2022}\natexlab{}.
\newblock \showarticletitle{Using deepspeed and megatron to train
  megatron-turing nlg 530b, a large-scale generative language model}.
\newblock \bibinfo{journal}{{\em arXiv preprint arXiv:2201.11990\/}}
  (\bibinfo{year}{2022}).
\newblock


\bibitem[\protect\citeauthoryear{Talebirad and Nadiri}{Talebirad and
  Nadiri}{2023}]%
        {talebirad2023multiagent}
\bibfield{author}{\bibinfo{person}{Yashar Talebirad} {and}
  \bibinfo{person}{Amirhossein Nadiri}.} \bibinfo{year}{2023}\natexlab{}.
\newblock \bibinfo{title}{Multi-Agent Collaboration: Harnessing the Power of
  Intelligent LLM Agents}.
\newblock   (\bibinfo{year}{2023}).
\newblock
\showeprint[arxiv]{cs.AI/2306.03314}


\bibitem[\protect\citeauthoryear{TGI}{TGI}{2023}]%
        {tgi}
TGI \bibinfo{year}{2023}\natexlab{}.
\newblock \bibinfo{title}{TGI}.
\newblock
  \bibinfo{howpublished}{\url{https://github.com/huggingface/text-generation-inference}}.
    (\bibinfo{year}{2023}).
\newblock
\newblock
\shownote{Accessed: 2023-10-21.}


\bibitem[\protect\citeauthoryear{Touvron, Martin, Stone, Albert, Almahairi,
  Babaei, Bashlykov, Batra, Bhargava, Bhosale, et~al\mbox{.}}{Touvron
  et~al\mbox{.}}{2023}]%
        {touvron2023llama}
\bibfield{author}{\bibinfo{person}{Hugo Touvron}, \bibinfo{person}{Louis
  Martin}, \bibinfo{person}{Kevin Stone}, \bibinfo{person}{Peter Albert},
  \bibinfo{person}{Amjad Almahairi}, \bibinfo{person}{Yasmine Babaei},
  \bibinfo{person}{Nikolay Bashlykov}, \bibinfo{person}{Soumya Batra},
  \bibinfo{person}{Prajjwal Bhargava}, \bibinfo{person}{Shruti Bhosale},
  {et~al\mbox{.}}} \bibinfo{year}{2023}\natexlab{}.
\newblock \showarticletitle{Llama 2: Open foundation and fine-tuned chat
  models}.
\newblock \bibinfo{journal}{{\em arXiv preprint arXiv:2307.09288\/}}
  (\bibinfo{year}{2023}).
\newblock


\bibitem[\protect\citeauthoryear{Van~Aken, Pavlo, Gordon, and Zhang}{Van~Aken
  et~al\mbox{.}}{2017}]%
        {van2017automatic}
\bibfield{author}{\bibinfo{person}{Dana Van~Aken}, \bibinfo{person}{Andrew
  Pavlo}, \bibinfo{person}{Geoffrey~J Gordon}, {and} \bibinfo{person}{Bohan
  Zhang}.} \bibinfo{year}{2017}\natexlab{}.
\newblock \showarticletitle{Automatic database management system tuning through
  large-scale machine learning}. In \bibinfo{booktitle}{{\em Proceedings of the
  2017 ACM international conference on management of data}}.
  \bibinfo{pages}{1009--1024}.
\newblock


\bibitem[\protect\citeauthoryear{Venkataraman, Panda, Ousterhout, Armbrust,
  Ghodsi, Franklin, Recht, and Stoica}{Venkataraman et~al\mbox{.}}{2017}]%
        {venkataraman2017drizzle}
\bibfield{author}{\bibinfo{person}{Shivaram Venkataraman},
  \bibinfo{person}{Aurojit Panda}, \bibinfo{person}{Kay Ousterhout},
  \bibinfo{person}{Michael Armbrust}, \bibinfo{person}{Ali Ghodsi},
  \bibinfo{person}{Michael~J Franklin}, \bibinfo{person}{Benjamin Recht}, {and}
  \bibinfo{person}{Ion Stoica}.} \bibinfo{year}{2017}\natexlab{}.
\newblock \showarticletitle{Drizzle: Fast and adaptable stream processing at
  scale}. In \bibinfo{booktitle}{{\em Proceedings of the 26th Symposium on
  Operating Systems Principles}}. \bibinfo{pages}{374--389}.
\newblock


\bibitem[\protect\citeauthoryear{Venkataraman, Yang, Franklin, Recht, and
  Stoica}{Venkataraman et~al\mbox{.}}{2016}]%
        {venkataraman2016ernest}
\bibfield{author}{\bibinfo{person}{Shivaram Venkataraman},
  \bibinfo{person}{Zongheng Yang}, \bibinfo{person}{Michael Franklin},
  \bibinfo{person}{Benjamin Recht}, {and} \bibinfo{person}{Ion Stoica}.}
  \bibinfo{year}{2016}\natexlab{}.
\newblock \showarticletitle{Ernest: Efficient performance prediction for
  $\{$Large-Scale$\}$ advanced analytics}. In \bibinfo{booktitle}{{\em 13th
  USENIX Symposium on Networked Systems Design and Implementation (NSDI 16)}}.
  \bibinfo{pages}{363--378}.
\newblock


\bibitem[\protect\citeauthoryear{Wang, Yi, Guo, Jin, Xu, Li, Wang, Guo, Li, Xu,
  Yu, Yuan, Zou, Long, Cai, Li, Zhang, Mo, Gu, Jiang, Wei, and Xie}{Wang
  et~al\mbox{.}}{2021}]%
        {Milvus}
\bibfield{author}{\bibinfo{person}{Jianguo Wang}, \bibinfo{person}{Xiaomeng
  Yi}, \bibinfo{person}{Rentong Guo}, \bibinfo{person}{Hai Jin},
  \bibinfo{person}{Peng Xu}, \bibinfo{person}{Shengjun Li},
  \bibinfo{person}{Xiangyu Wang}, \bibinfo{person}{Xiangzhou Guo},
  \bibinfo{person}{Chengming Li}, \bibinfo{person}{Xiaohai Xu},
  \bibinfo{person}{Kun Yu}, \bibinfo{person}{Yuxing Yuan},
  \bibinfo{person}{Yinghao Zou}, \bibinfo{person}{Jiquan Long},
  \bibinfo{person}{Yudong Cai}, \bibinfo{person}{Zhenxiang Li},
  \bibinfo{person}{Zhifeng Zhang}, \bibinfo{person}{Yihua Mo},
  \bibinfo{person}{Jun Gu}, \bibinfo{person}{Ruiyi Jiang}, \bibinfo{person}{Yi
  Wei}, {and} \bibinfo{person}{Charles Xie}.} \bibinfo{year}{2021}\natexlab{}.
\newblock \showarticletitle{{Milvus: A Purpose-Built Vector Data Management
  System}}. In \bibinfo{booktitle}{{\em Proceedings of the ACM International
  Conference on Management of Data (SIGMOD)}}. \bibinfo{pages}{2614--2627}.
\newblock


\bibitem[\protect\citeauthoryear{Wang, Jia, Zhang, Zhang, Fu, Ng, and
  Wang}{Wang et~al\mbox{.}}{2023}]%
        {wang2023gemini}
\bibfield{author}{\bibinfo{person}{Zhuang Wang}, \bibinfo{person}{Zhen Jia},
  \bibinfo{person}{Shuai Zhang}, \bibinfo{person}{Zhen Zhang},
  \bibinfo{person}{Mason Fu}, \bibinfo{person}{TS~Eugene Ng}, {and}
  \bibinfo{person}{Yida Wang}.} \bibinfo{year}{2023}\natexlab{}.
\newblock \showarticletitle{Gemini: Fast failure recovery in distributed
  training with in-memory checkpoints}.
\newblock  (\bibinfo{year}{2023}).
\newblock


\bibitem[\protect\citeauthoryear{Wei, Tay, Bommasani, Raffel, Zoph, Borgeaud,
  Yogatama, Bosma, Zhou, Metzler, et~al\mbox{.}}{Wei et~al\mbox{.}}{2022}]%
        {wei2022emergent}
\bibfield{author}{\bibinfo{person}{Jason Wei}, \bibinfo{person}{Yi Tay},
  \bibinfo{person}{Rishi Bommasani}, \bibinfo{person}{Colin Raffel},
  \bibinfo{person}{Barret Zoph}, \bibinfo{person}{Sebastian Borgeaud},
  \bibinfo{person}{Dani Yogatama}, \bibinfo{person}{Maarten Bosma},
  \bibinfo{person}{Denny Zhou}, \bibinfo{person}{Donald Metzler},
  {et~al\mbox{.}}} \bibinfo{year}{2022}\natexlab{}.
\newblock \showarticletitle{Emergent abilities of large language models}.
\newblock \bibinfo{journal}{{\em arXiv preprint arXiv:2206.07682\/}}
  (\bibinfo{year}{2022}).
\newblock


\bibitem[\protect\citeauthoryear{Wolf, Debut, Sanh, Chaumond, Delangue, Moi,
  Cistac, Rault, Louf, Funtowicz, et~al\mbox{.}}{Wolf et~al\mbox{.}}{2019}]%
        {wolf2019huggingface}
\bibfield{author}{\bibinfo{person}{Thomas Wolf}, \bibinfo{person}{Lysandre
  Debut}, \bibinfo{person}{Victor Sanh}, \bibinfo{person}{Julien Chaumond},
  \bibinfo{person}{Clement Delangue}, \bibinfo{person}{Anthony Moi},
  \bibinfo{person}{Pierric Cistac}, \bibinfo{person}{Tim Rault},
  \bibinfo{person}{R{\'e}mi Louf}, \bibinfo{person}{Morgan Funtowicz},
  {et~al\mbox{.}}} \bibinfo{year}{2019}\natexlab{}.
\newblock \showarticletitle{Huggingface's transformers: State-of-the-art
  natural language processing}.
\newblock \bibinfo{journal}{{\em arXiv preprint arXiv:1910.03771\/}}
  (\bibinfo{year}{2019}).
\newblock


\bibitem[\protect\citeauthoryear{Wu, Bansal, Zhang, Wu, Zhang, Zhu, Li, Jiang,
  Zhang, and Wang}{Wu et~al\mbox{.}}{2023}]%
        {wu2023autogen}
\bibfield{author}{\bibinfo{person}{Qingyun Wu}, \bibinfo{person}{Gagan Bansal},
  \bibinfo{person}{Jieyu Zhang}, \bibinfo{person}{Yiran Wu},
  \bibinfo{person}{Shaokun Zhang}, \bibinfo{person}{Erkang Zhu},
  \bibinfo{person}{Beibin Li}, \bibinfo{person}{Li Jiang},
  \bibinfo{person}{Xiaoyun Zhang}, {and} \bibinfo{person}{Chi Wang}.}
  \bibinfo{year}{2023}\natexlab{}.
\newblock \showarticletitle{AutoGen: Enabling Next-Gen LLM Applications via
  Multi-Agent Conversation Framework}.
\newblock
\showeprint[arxiv]{cs.AI/2308.08155}


\bibitem[\protect\citeauthoryear{Wu, Lentz, Zhuo, and Lu}{Wu
  et~al\mbox{.}}{2022}]%
        {wu2022serving}
\bibfield{author}{\bibinfo{person}{Yongji Wu}, \bibinfo{person}{Matthew Lentz},
  \bibinfo{person}{Danyang Zhuo}, {and} \bibinfo{person}{Yao Lu}.}
  \bibinfo{year}{2022}\natexlab{}.
\newblock \showarticletitle{Serving and Optimizing Machine Learning Workflows
  on Heterogeneous Infrastructures}.
\newblock \bibinfo{journal}{{\em arXiv preprint arXiv:2205.04713\/}}
  (\bibinfo{year}{2022}).
\newblock


\bibitem[\protect\citeauthoryear{Yu, Jeong, Kim, Kim, and Chun}{Yu
  et~al\mbox{.}}{2022}]%
        {yu2022orca}
\bibfield{author}{\bibinfo{person}{Gyeong-In Yu}, \bibinfo{person}{Joo~Seong
  Jeong}, \bibinfo{person}{Geon-Woo Kim}, \bibinfo{person}{Soojeong Kim}, {and}
  \bibinfo{person}{Byung-Gon Chun}.} \bibinfo{year}{2022}\natexlab{}.
\newblock \showarticletitle{Orca: A distributed serving system for
  $\{$Transformer-Based$\}$ generative models}. In \bibinfo{booktitle}{{\em
  16th USENIX Symposium on Operating Systems Design and Implementation (OSDI
  22)}}. \bibinfo{pages}{521--538}.
\newblock


\bibitem[\protect\citeauthoryear{Yuan, He, Davis, Zhang, Dao, Chen, Liang, Re,
  and Zhang}{Yuan et~al\mbox{.}}{2022}]%
        {yuan2022decentralized}
\bibfield{author}{\bibinfo{person}{Binhang Yuan}, \bibinfo{person}{Yongjun He},
  \bibinfo{person}{Jared Davis}, \bibinfo{person}{Tianyi Zhang},
  \bibinfo{person}{Tri Dao}, \bibinfo{person}{Beidi Chen},
  \bibinfo{person}{Percy~S Liang}, \bibinfo{person}{Christopher Re}, {and}
  \bibinfo{person}{Ce Zhang}.} \bibinfo{year}{2022}\natexlab{}.
\newblock \showarticletitle{Decentralized training of foundation models in
  heterogeneous environments}.
\newblock \bibinfo{journal}{{\em Advances in Neural Information Processing
  Systems\/}}  \bibinfo{volume}{35} (\bibinfo{year}{2022}),
  \bibinfo{pages}{25464--25477}.
\newblock


\bibitem[\protect\citeauthoryear{Zaharia, Das, Li, Hunter, Shenker, and
  Stoica}{Zaharia et~al\mbox{.}}{2013}]%
        {zaharia2013discretized}
\bibfield{author}{\bibinfo{person}{Matei Zaharia}, \bibinfo{person}{Tathagata
  Das}, \bibinfo{person}{Haoyuan Li}, \bibinfo{person}{Timothy Hunter},
  \bibinfo{person}{Scott Shenker}, {and} \bibinfo{person}{Ion Stoica}.}
  \bibinfo{year}{2013}\natexlab{}.
\newblock \showarticletitle{Discretized streams: Fault-tolerant streaming
  computation at scale}. In \bibinfo{booktitle}{{\em Proceedings of the
  twenty-fourth ACM symposium on operating systems principles}}.
  \bibinfo{pages}{423--438}.
\newblock


\bibitem[\protect\citeauthoryear{Zhang, Roller, Goyal, Artetxe, Chen, Chen,
  Dewan, Diab, Li, Lin, et~al\mbox{.}}{Zhang et~al\mbox{.}}{2022}]%
        {zhang2022opt}
\bibfield{author}{\bibinfo{person}{Susan Zhang}, \bibinfo{person}{Stephen
  Roller}, \bibinfo{person}{Naman Goyal}, \bibinfo{person}{Mikel Artetxe},
  \bibinfo{person}{Moya Chen}, \bibinfo{person}{Shuohui Chen},
  \bibinfo{person}{Christopher Dewan}, \bibinfo{person}{Mona Diab},
  \bibinfo{person}{Xian Li}, \bibinfo{person}{Xi~Victoria Lin},
  {et~al\mbox{.}}} \bibinfo{year}{2022}\natexlab{}.
\newblock \showarticletitle{Opt: Open pre-trained transformer language models}.
\newblock \bibinfo{journal}{{\em arXiv preprint arXiv:2205.01068\/}}
  (\bibinfo{year}{2022}).
\newblock


\end{thebibliography}

\end{document}